\newcommand{\ergs}{erg~s$^{-1}$}
\documentclass[useAMS,usenatbib]{mn2e}
\usepackage{amssymb}
\usepackage{graphicx}
\usepackage{caption}
\usepackage{mwe}
\usepackage{pdflscape}
\usepackage{capt-of}
\usepackage{multirow}
\usepackage{color}

\begin{document}

\title[Long Term MIR Decline in Galaxies]{Long-Term Decline of the Mid-Infrared 
Emission of Normal Galaxies: Dust Echo of Tidal Disruption Flare?}
\author[T.G. Wang et al.]{Tinggui Wang$^{1,2}$\thanks{E-mail:twang@ustc.edu.cn}, 
Lin Yan$^{3,4}$, 
Liming Dou$^{5,6}$, Ning Jiang$^{1,2}$, Zhenfeng Sheng$^{1,2}$, 
\newauthor and Chenwei Yang$^{1,2}$\\ 
$^{1}$CAS Key Laboratory for Research in Galaxies and Cosmology, University 
of Science and Technology of China, Hefei, Anhui, 230026, China \\
$^{2}$ School of Astronomy and Space Science, University of Science and Technology of China, 
Hefei, Anhui, 230026, China \\
$^{3}$ Caltech Optical Observatories, Cahill Center for Astronomy and Astrophysics, 
California Institute of Technology, Pasadena, CA 91125, USA;\\ 
$^{4}$Infrared Processing and Analysis Center, California Institute of Technology, 
Pasadena, CA 91125, USA 0000-0003-1710-9339 \\
$^{5}$ Center for Astrophysics, Guangzhou University, Guangzhou 510006, China; \\ 
$^{6}$ Astronomy Science and Technology Research Laboratory of Department of Education of Guangdong Province, Guangzhou 510006, China}
\date{Accepted 2018 Februray 9. Received 2018 Feburary 2; in original form 2017 May 7}

\pagerange{\pageref{firstpage}--\pageref{lastpage}} \pubyear{2002}

\maketitle

\label{firstpage}
\begin{abstract}

We report the discovery of a sample of 19 low redshift ($z<0.22$) spectroscopically  non-Seyfert 
galaxies that show slow declining mid-infrared (MIR) light-curves (LCs), similar to those of tidal disruption 
event (TDE) candidates with extreme coronal lines.  Two sources also showed a relatively fast rising 
MIR LCs. They consist of 61\% sample of the WISE MIR variable non-Seyfert galaxies with SDSS spectra. 
In a comparison sample of optically selected Seyfert galaxies, the fraction of sources with such a LC is only 
15\%. After rejecting 5 plausible obscured Seyfert galaxies with red MIR colours, remaining 14 objects are 
studied in detail in this paper. We fit the declining part of LC with an exponential law, and the decay time is 
typically one year.  The observed peak MIR luminosities ($\nu L_\nu$) after subtracting host galaxies are in 
the range of a few $10^{42}$ to $10^{44}$ erg~s$^{-1}$ with a median of $5\times10^{43}$ erg~s$^{-1}$ in 
the $W2$ band. The black hole masses distribute in a wide range with more than half in between 
$10^7$ to $10^8$~$M_{\sun}$, but significantly different from that of optical/UV selected TDEs.  
Furthermore, MIR luminosities are correlated with black hole masses, the stellar 
mass or luminosity of their host bulges. Most galaxies in the sample are red and luminous with an absolute 
magnitude of $r$ between -20 to -23. We estimate the rate of event about $10^{-4}$ gal$^{-1}$~yr$^{-1}$ 
among luminous red galaxies. We discuss several possibilities for the variable infrared sources, and conclude 
that most likely, they are caused by short sporadic fueling to the supermassive black holes via either the 
instability of accretion flows or tidal disruption of stars.  

\end{abstract}

\section[]{Introduction}

A star passing too close to the supermassive black hole (SMBH) in the galactic nucleus 
will be torn apart by the tidal force of the hole \citep{Hills75}. About half of the 
debris is accreted onto the black hole, producing a strong UV to soft X-ray flare 
\citep{Rees88,Phinney89,Komossa15}. The flare rises rather fast on a time scale of 
about a month \citep{Gezari09,Gezari12}, and fades approximately in a power-law 
form \citep{Komossa99, Gezari15, Brown16, vanVelzen16, Brown17}.  
Transient broad emission lines have been detected recently in a significant fraction 
of known tidal disruption events (TDEs) or TDE candidates although their origin has 
not yet fully understood \citep{Komossa08,vanVelzen11,Wang11, Gezari12, Arcavi14}. 
Up to now, several dozens of tidal disruption events have been reported, and the 
rate of events is broadly consistent with theoretical predictions at about 
$10^{-5}$ to $10^{-4}$ gal$^{-1}$~yr$^{-1}$ \citep{Wang04,Donley02,Esquej08, 
Wang12, Velzen14, Holoien16a, Stone16}. These events provide a unique opportunity to 
probe SMBHs in quiescent galaxies at the low mass end, below a few $10^6$ 
M$_\odot$, or at redshifts beyond 0.1, where it is difficult to probe SMBHs 
using other techniques. 

Illuminated by the transient UV and X-ray emission, gas surrounding the black hole 
is photo-ionized and produce emission lines \citep{Ulmer99}. \citet{Komossa08} 
reported a transient extreme coronal line emitter (ECLE) in J0952+2143 and 
suggested that it might be caused by TDE. A moderate-amplitude optical flare was 
observed serendipitously in the optical survey carried out in the Lincoln Near-Earth 
Asteroid Research asteroid survey (LINEAR)\citep{Palaversa16}. The light-curve covers 
the rising phase, and the detection of large amplitude UV variability is consistent 
with a TDE. Transients with extreme variable coronal lines were detected in three 
other star 
forming galaxies \citep{Wang12,Yang13}. Three of four ECLEs also display transient broad 
emission lines, including the first reported example of strong blue-shifted very broad 
HeII lines associated with TDE candidates \citep{Wang11,Wang12}. \citet{Yang13} also 
observed brightening of [O {\sc iii}] emission lines 5-7 years after the discovery of the high 
ionization lines in two ECLEs. The [O III] doublet line ratios are unusual, substantially higher 
(or lower) than the canonical values, suggesting very thick warm gas surrounding the 
central black holes.
  
Dust within a few parsecs of the black hole will absorb UV flux and re-emit in the 
infrared. \citet{Lu16} considered a model of spherically distributed dust with a sky 
covering factor 0.1 to 1.0 within 1 pc of the black hole, and predicted  an infrared 
luminosity of order of $10^{42}$ to $10^{43}$ erg~s$^{-1}$, lasting for several years 
for a typical observed TDE. \citet{Jiang16}  found an infrared flare in both $W1$ and 
$W2$ bands of Wide Infrared Sky Explorer (WISE, \citet{Wright10}) around 36 days after 
the discovery of the optical/X-ray flare, or a hundred days after the peak, for the 
nearest known TDE ASASSN-14li. The infrared luminosity is nearly two orders of magnitude 
lower than the model prediction, suggesting either small covering factor or small amount 
dust in the nucleus.  \citet{vanVelzen16} detected 3 out of five bright TDEs searched in 
the 3.4$\mu$m WISE band, and they also argued that dust covering factor must be small. 
While \citet{Dou16} examined MIR light curves obtained by WISE for four ECLEs, and found 
that they all displayed long-term fading in both $W1$ and $W2$ bands 4-8 years after the 
initial detections of transient coronal lines. The infrared luminosities were two orders 
of magnitude higher than that in ASASSN-14li.

In view of these infrared echoes detected in TDE candidates. We have examined systematically 
the WISE light curves of optically normal galaxies, in hope to find new infrared echoes 
of TDEs. This is important as infrared emission is subject less to the dust extinction 
than optical and UV light. Hitherto, only the jetted TDE SWIFT J1644+57 was known to occur 
on the partially obscured nucleus ($E(B-V)=2.0$), which showed variable infrared 
emission \citep{Levan16}. The paper is arranged as follows. We present the sample in 
\S 2. The light curves are analysed in \S 3. Simple statistics are given in \S 4. We discuss 
the implication in \S 5. Throughout the paper, we will adopt a cosmology of ($H_0,\,\Omega_M,
\,\Omega_\Lambda)$=(70 km~s$^{−1}$~Mpc$^{−1}$, 0.3, 0.7)
 
\section[]{Infrared Flare non-Seyfert Galaxies}

\subsection{Classification of ALLWISE infrared variable sources with SDSS spectra} 

We cull mid-infrared variable sources from ALLWISE catalogue, that includes the Cryogenic 
WISE all sky survey \citep{Wright10}, WISE 3-band Survey and post-Cryogenic NEOWISE 
survey \citep{Mainzer11}, using the flags provided by the standard pipeline. The selection 
criteria are following: (1) point sources with $ext\_flag=0$ to avoid additional photometric 
errors that unaccounted by error model; (2) $var\_flg\geq 7$ in at least one band; (3) 
no source contamination in W1 and W2 band $cc\_flags="00??"$. The $var\_flg$ is a 
combination of two components: the implausibility of the null hypothesis that the source 
is not variable, and the significance of band to band correlations. Sources with 
$var\_flg > 6$ can generally be considered as having a high reliability of being 
variable. Noting the variability flag is based on the data taken between Dec. 14, 2009 
and Feb. 1, 2011 (ALLWISE data) only. This yields a list of total 96,856 sources. From their 
distribution on the sky, it is evident that most of them are Galactic objects. It should 
be noted that the above criteria still miss a large fraction of the mid-infrared sources 
variable on long-term as in the case for ECLEs, where only one of four objects meets above 
criteria, while all of them show long-term variations, as presented in \citet{Dou16}. 

Next, we cross-correlate the above list with the SDSS spectroscopic catalogue. It results 
in 666 matched SDSS spectra for 526 objects within 3 arcsecs matching radius. Of these 
526 sources, 360 are classified by SDSS pipeline as "QSO", 99 as "GALAXY", 31 as "STAR" 
and the remaining 36 unclassified\footnote{For a source with more than one spectrum, we 
choose the one with the best signal to noise ratio.}. Note that "QSO" includes all subtypes 
of broad-line AGN and BL Lac objects. We extracted SDSS spectra from SDSS DR 12 
\citep{Alam15} archive and checked the spectra visually. In this way, we find that  
about half of objects classified as 'GALAXY' by the SDSS pipeline are actually BL Lacs 
(featureless continuum and strong radio emission) or broad line AGNs, and we correct 
for this, and also make a classification for 36 objects that were not classified by SDSS pipeline.

We also add a blazar subclass, which includes BL Lac and flat radio quasars. In our 
sample, 245 were detected in the FIRST survey (version 2014 December 17; \citep{Becker95}) 
with a median radio flux of 86.45~mJy. The large median radio flux is due to the fact that 
the FIRST detected sub-sample is dominated by radio strong BL-Lac objects and flat 
spectral radio quasars (FSRQ). Note that about 10\% percent sources fall outside the FIRST 
survey footprint, in which we also check the NVSS catalogue \citep{Condon98}. 27 additional 
sources are found in the NVSS catalogue. Figure \ref{radio} shows the distribution  
radio flux. The distribution for the subsample of "GALAXY" classified by SDSS spectroscopic 
pipeline is indistinguishable from the whole sample according to K-S test. According 
to our visual examinations, most of them are BL Lac objects with a prominent featureless 
continuum. Although we do not have radio spectral index information, most radio-strong 
($f_{\mathrm 1.4GHz}>3.0$ mJy) broad-line objects in this sample show systematically weaker 
emission lines and shorter term infrared variability at a given luminosity. Therefore, they 
are most likely blazars. We put all broad-line AGNs with $f_{\mathrm 1.4GHz}>3.0$ mJy and all 
sources with featureless blue continuum regardless of its radio flux into the subclass of blazars.

To further classify narrow emission line galaxies, we extract emission line flux from 
Portsmouth database \citep{Thomas13}. In the first step, a spectrum was decomposed into 
starlight and nebular emission lines using the publicly available codes Penalized PiXel 
Fitting (pPXF, Cappellari \& Emsellem 2004). Next, they used Gas and Absorption Line 
Fitting code (GANDALF v1.5; Sarzi et al. 2006) to derive emission line parameters by 
fitting the nebular component with Gaussian templates. The first step is essential as 
some emission lines, especially H$\beta$, are severely affected by the stellar absorption 
lines. Each emission line was modeled with one gausssian. We examine the 
optical spectrum and find that our objects usually show narrow and symmetrical emission 
lines, and a single gaussian should be sufficient. Objects show only weak ($W_{H\alpha}<2$\AA) 
or no detectable H$\alpha$ line ($SNR<5.0$) are classified as pure absorption line galaxies. 

\begin{figure*}
\centering
\begin{minipage}[c]{\textwidth}
  \centering
     \includegraphics[width=6.0in]{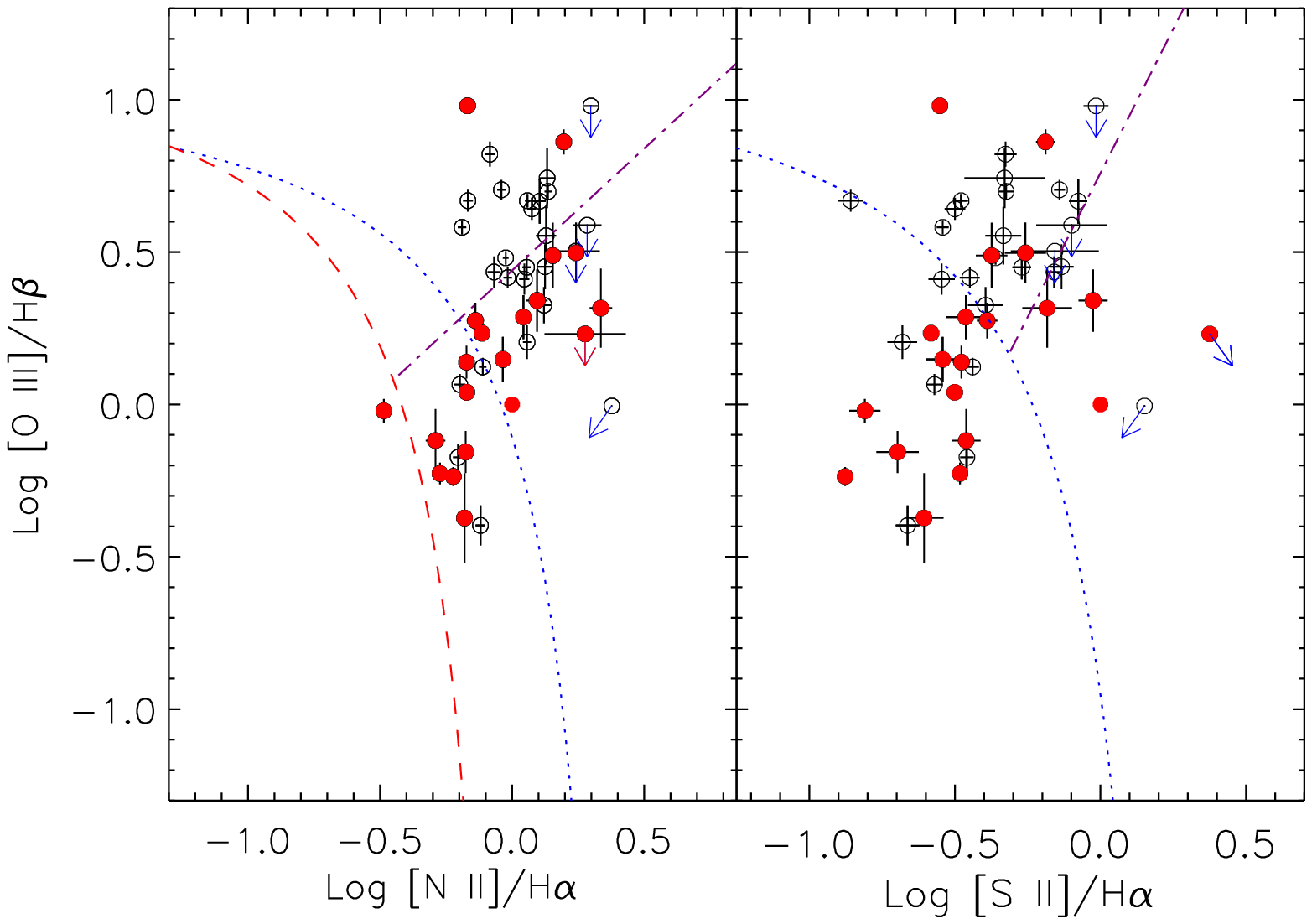}
     \caption{The BPT diagram of WISE variable narrow-emission-line galaxies. 
The red-dashed curve and dotted curve are the demarcation lines defined by Kewley 
et al. (2006) for normal and extreme star-forming galaxies, respectively, and 
dot-dashed line separates LINERs from Seyfert galaxies. The red filled circles are 
the sources with a flare-like light curve.}
     \label{bpt}
\end{minipage}
\end{figure*}

\begin{figure*}
\centering
\begin{minipage}[c]{\textwidth}
  \centering
     \includegraphics[width=6.0in]{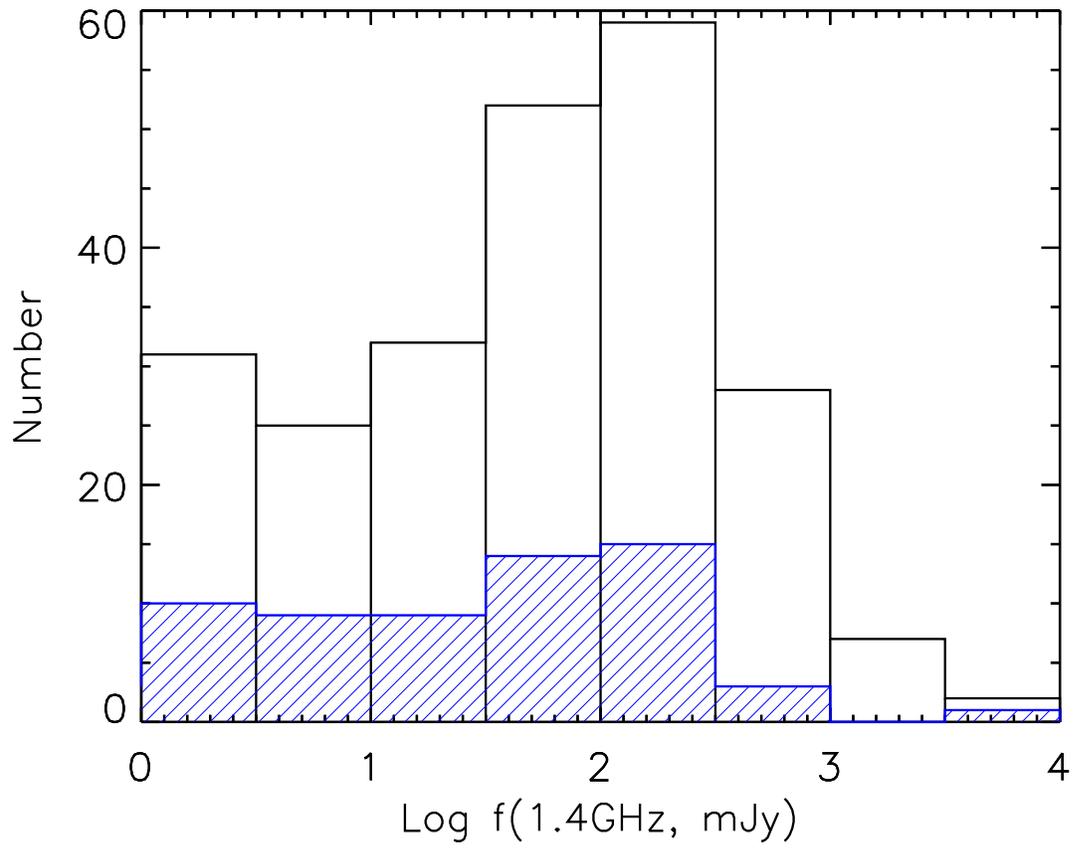}
     \caption{The distributions of radio flux for the whole sample (open) and the sub-sample of 
'GALAXY' (hatched) classified by the SDSS pipeline. Many of these galaxies are actually blazars 
according to their optical spectra. }
     \label{radio}
\end{minipage}
\end{figure*}

Using emission line ratios, we can further classify emission line galaxies into different 
subclasses according to their location on Baldwin, Philips and Terlevich (BPT) diagrams 
\citep{Kewley06}. Among 45 galaxies (Figure \ref{bpt}) with reliable measurements of 
H$\alpha$ and H$\beta$ or [N {\sc ii}], 15 are classified as Seyfert 2 galaxies, 11 as 
star forming galaxies and 19 as Low Ionization Nuclear Emission Region (LINERs). These 
objects with non-detection of [O {\sc iii}] is considered as star-forming galaxies or 
absorption line according to the EWs of H$\alpha$ line ($W_{H\alpha}<2$\AA~ for absorption 
line galaxies). It should be noted that a large fraction of LINERs may be actually ionized 
by evolved star or radiative shocks rather than by an AGN as LINER-like spectra often appears 
in off-nucleus of normal galaxies as well (e.g., Cid Fernandes et al. 2011; also Zhang 
et al. 2017).

With all above vetting, we find the number of sources in each class as follows: 238 
radio-weak broad-line AGNs, 218 blazars, 24 Seyfert 2 galaxies, 31 normal galaxies, 
including 1 absorption line galaxy, 5 star-forming galaxies and 8 LINERs, and 14 
stars. Blazar is the second largest class after radio weak broad-line AGNs despite 
they are much less common among AGNs, reflecting that they are more variable than 
other subclasses of AGNs. In the following sections, we will mainly focus on these 
non-Seyfert galaxies. 

\subsection[]{Selection of MIR Flare Sources}

We extracted MIR light-curves from the public WISE archive for all 526 sources. The 
WISE survey is good for detecting short time scale variabilities on time scale of hours, 
because the same piece of sky is usually scanned more than 12 times within two days, 
depending on the sky position. In order to increase signal to noise ratio, following 
\citet{Jiang16}, we rebinned light curves so that the photometry within two days is 
co-added. In practice, we fit a constant to the infrared magnitude within two days by 
minimizing $\chi^2$. To account for potential other errors in the measurements, we also 
add a term $\sigma_{unknown}$ to the error so $\sigma^2=\sigma_{wise}^2+\sigma_{unknown}^2$ 
to ensure the $\chi^2/dof\simeq 1$. Initially we set $\sigma_{unknown}$ to zero, and then 
iteratively adjust $\sigma_{unknown}$ until $\chi^2/dof\simeq 1$. We found a typical 
$\sigma_{unknown}\sim 0.06$, which is comparable to the uncertainty given by the 
pipeline that listed in WISE All Sky survey documents \footnote{http://wise2.ipac.caltech.edu/docs/release/allsky/}. 
In addition, we supplement additional NEOWISE photometries for these 
256 objects, extending time coverage to 2300 days.

Our purpose is to search for TDE candidates with mid-IR echoes. So, we focus only on the 
subset of 31 normal galaxies. In other words, their optical spectra and radio photometries 
have little signs of Seyfert type activities. In addition, we require that the mid-IR 
light curve either declines monotonically or displays only one peak, to distinguish the 
variability of Seyfert galaxies, which usually show up certain stochastic characteristics. 
Mathematically, a peak is the data point in the light curve that the flux rises before it 
and declines after that. For each data point, we decide whether it is a local peak or not, 
by searching backward for the evidence of a significant rising and forward for a significant 
decline. In practice, the fluctuations due to measurement errors will cause false 'rising' 
or 'falling' in the light curve. To tackle this, we define 2$\sigma$ increase or decrease 
over the previous data points as a significant rising or falling to account for the 
uncertainties of the measurements. We carry out Monte-Carlo simulations to quantify, under 
2$\sigma$ threshold, how many single 
peaked light-curves are mis-identified with two or more peaks, and how many cases 
changed in opposite direction due to measurement errors. For each object, we 
simulate 100 light curves in W1 band and W2 bands. In each simulation,  we add 
randomly gaussian errors with amplitude as measured one to the observed light 
curve. Then, we measure the number of peaks in the light curve exactly in the 
same way as had carried on real light curve. We find that about 7.5\% single 
peaked light curves are misclassified as two or more peaks, and about 9.1\% 
light curves with two or more are erroneously as one peaked objects for the 
sample of galaxies.  In passing, we note that we use above non-parameter 
statistics to describe the light curve rather than using a light curve 
fitting because there is no prior knowledge on the form of the MIR light curve, 
which certainly depends on the dust distribution as well the form of UV light-curve. 

\begin{figure*}
\centering
\begin{minipage}[c]{\textwidth}
  \centering
    \includegraphics[width=6.0in]{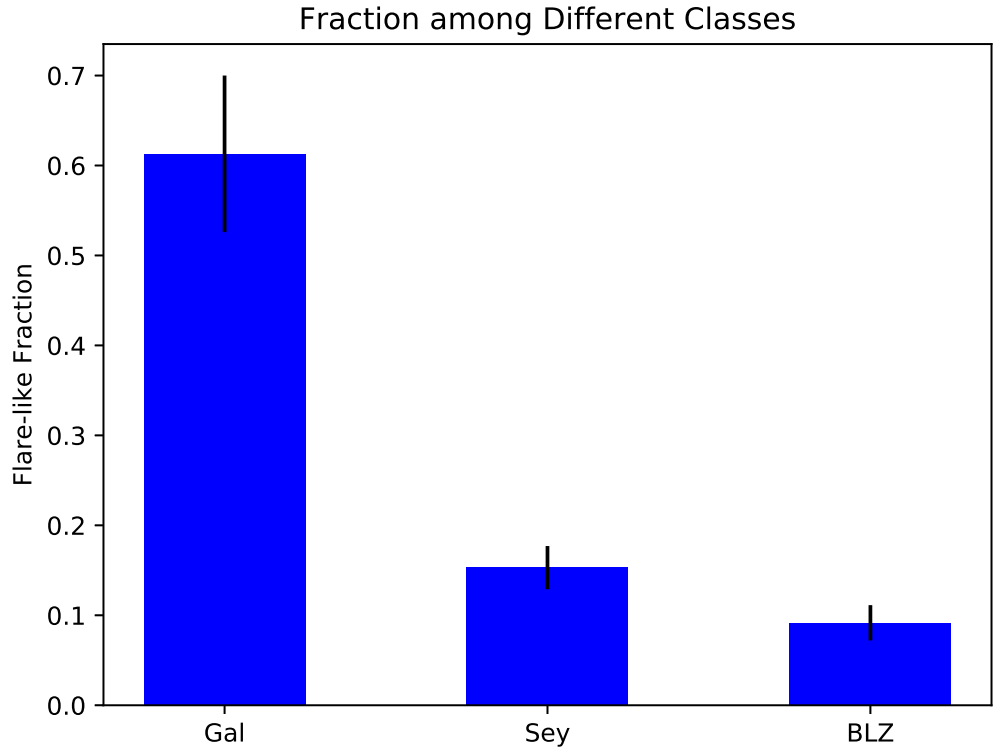}
    \caption{Fraction of flare-like sources among different subsamples of WISE variability-flag 
selected objects. For X-labels, Non-Seyfert galaxies are abbreviated to 'Gal', and Seyfert galaxies are shortened as 'Sey' and 
Blazars as 'BLZ'.}
    \label{fracflare}
\end{minipage}
\end{figure*}

Of 31 sources, 19 meet these criteria. This is 61.3\%. For comparison, we apply the 
same screening to the subsample of 229 Type 1 and Type 2 Seyfert galaxies in 
the redshift $z<0.25$, similar to those of galaxy sample, and find 35 (15.2\%) objects with 
flare-like light curves. The Seyfert galaxies in the comparison sample has 
similar W2 magnitude (with a median offset of only 0.12 magnitudes to the 
non-Seyfert galaxy sample).  It is clear that the fraction showing flare-like light 
curves is much higher in the sample of MIR variable normal galaxies than in 
the sample of Seyfert galaxies. The difference is statistically significant at 
2.5 $\sigma$. We make a similar analysis of light curves of the blazar 
subsample. Since more than half blazars in this sample do not have reliable 
redshifts, we do not apply any cut in the redshift. We find that 20 of 218 blazars 
show flare-like light curves. The fraction (9.2\%) is significantly smaller than 
that in non-Seyfert galaxies. These results are summarized in Figure 
\ref{fracflare}. 

In the search for a microlensing event in a light curve ($x_i=x(t_i)$) , 
Price-Whelan et al. (2014) used five statistical measures of variability compiled 
by Shin et al (2009) and also introduced von Neumann ratio (also known as 
Durbin-Watson statistics, von Neumann 1941; Durbin \& Watson 1950). Here, we adopt 
two of these six parameters, the asymmetric index and von Neumann ratio to 
characterize the light curve. The asymmetric index ($J$) is defined as follows:
\begin{equation}
\delta_i=\sqrt{\frac{N}{N-1}}\frac{x_i-\mu}{e_i}
\end{equation}
\begin{equation}
J=\sum_{i}^{N-1}sign(\delta_i\delta_{i+1})\sqrt{|\delta_i\delta_{i+1}|}
\end{equation}
where $e_i$ is the error of $x_i$, $\mu$ is the mean of $x_i$, and the $sign$ returns 
the sign of the variable. For a non-variable source, $J$ will be close to zero; for a 
well sampled light-curve of a strong variable source, $J$ should be large. However, if 
the cadence is comparable to the characteristic variability time scale, then $J$ can be 
a large negative, positive or close to zero. An example is a periodic light curve, and 
the observations happen to sample maxima and minima. 
  
The von Neumann ratio is 
\begin{equation}
\eta=\frac{\sum_i^{N-1}(x_{i+1}-x_i)^2/(N-1)}{\sigma^2}
\end{equation}
where $\sigma$ is the variance of the time series. A small $\eta$ implies a 
strong positive serial correlation between the successive data points. Applying 
to our case, $\eta$ is small for a continuously declining or rising light curve and 
large for random fluctuations. So we expect that a long term flare-like light-curve 
will have a smaller $\eta$ than stochastic variations as seen in AGNs, which can be 
described as damping random walks (DRW, e.g., Kelly et al. 2009). 

\begin{figure*}
\centering
\begin{minipage}[c]{\textwidth}
  \centering
    \includegraphics[width=6.0in]{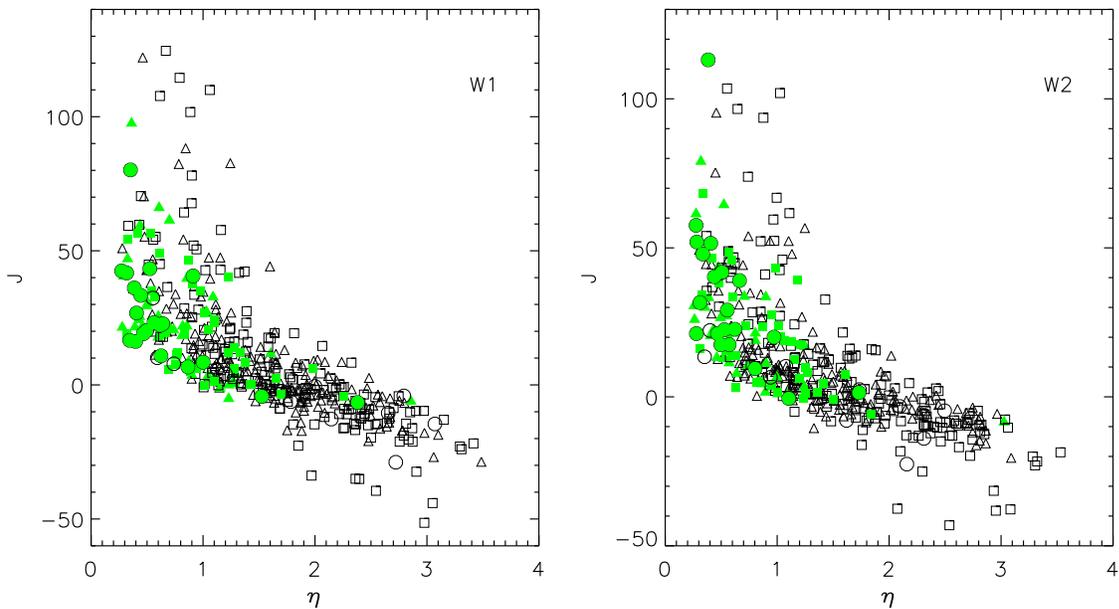}
    \caption{
The asymmetric parameter $J$ versus von Neumann ratio $\eta$ for non-Seyfert galaxies 
(circles), Seyfert galaxies (triangles) and Blazars (squares). Filled green symbols 
represent objects with a flare-like light-curve.  
The $W1$ band is shown on the left panel and $W2$ on the right panel.  }
    \label{vonneumann}
\end{minipage}
\end{figure*}

We calculate $J$ and $\eta$ for 31 spectroscopic non-Seyfert galaxies, 229 Seyfert 
galaxies at redshift $z<0.25$ and 218 blazars. We show $J$ versus $\eta$ for each 
subsample in Figure \ref{vonneumann}. First, most objects have large $J$, suggesting 
that typical rising or falling trend on time scale scales longer than the cadence of 
order a half year. In general, blazars are more likely with a large positive or 
negative $J$. The latter is likely caused by large-amplitude short-term variability. 
Second, non-Seyfert galaxies have a larger fraction of objects with small $\eta$. We 
perform the Kolmogorov–Smirnov test for whether two distributions are drawn from the 
same parent population, and find that Seyfert galaxies and blazars show marginally 
different $\eta$ distributions with $D=0.145, 0.185$, $P_{KS}=0.02,0.001$ in $W1$ and 
$W2$ bands, respectively. While non-Seyfert galaxies are significantly different from 
either Seyfert galaxies ($D=0.408,0.410$ and $P_{KS}=1\times10^{-4}, 1\times10^{-4}$) 
or blazars ($D=0.443,0.465$ and $P=2\times 10^{-5},6\times10^{-6}$) in $W1$ and $W2$, 
respectively. Finally, objects with one-peak light-curve on average have smaller 
$\eta$ and a positive $J$, qualitatively consistent with our intuition. But there 
are some exceptions. An example of the latter is J141036.81+265425.0, where $J$ is 
small negative and $\eta$ is large ($>2$). This is caused poor sampling of the 
flare, particularly, only one data point is notably above the quiescent level in the 
$W1$-light curve. These exceptions indicates that $\eta$ and $J$ parameters are not 
as robust as our single-peak parameter in the application to the case of sparse sampling.

As a final check, we examine the infrared colour $W1-W2$ in the lowest flux level 
to further screen potential obscured AGNs. According to \citet{Assef10}, $W1-W2>0.8$ 
can select pure AGNs very effectively. Thus, we reject additional 5 sources with 
$W1-W2>0.75$. The final sample consists of 14 objects.   Many galaxies 
locate on the starburst and LINER regime with W1-W2$\sim$0.5. 
The basic properties of these galaxies are summarized in Table \ref{table1}. 
Figure \ref{lcsp} shows the light curves and SDSS spectra of all these sources. 
The sample covers the redshift range from 0.0365 to 0.2164. 

It is intriguing that in the sample, only two sources show also the uprising phase. 
This suggests that the light curve must be very asymmetry, with a steep rise and a 
slow decay, so one has a higher chance of detection the falling part of the light 
curve than the rising part in the ALLWISE data if flare lasts longer than the period 
of WISE survey. The light-curve of these two sources seem to confirm that the 
rising time is much shorter than the decline time although we cannot give specific 
value due to poor sampling. The selection of WISE highly variable sources in the 
first two year survey may bias towards the sources in the rising phase because 
sources in the rising phase usually demonstrate larger variability amplitude, thus 
less likely be missed, than the sources on the falling part of light curve.  

\section[]{Properties of the Flares}

\subsection{Characterizing Infrared Flares}

We fit the declining part of the MIR light curves based on an exponential law 
to estimate the characteristic time scales (refer as decline time in the following 
sections). In ASASSN-14li, the reprocessed MIR emission is detected significantly 
only at one epoch in 2 bands, so we cannot learn much about the form of its 
light-curve, except the rising time scale is about 4-6 months after 
the UV flare. However, the MIR LCs of TDE candidates with ECLs has continuously 
faded for over 5 years from 5-9 years after their optical discoveries, and their 
MIR LCs can be approximately described by an exponential law or a power-law 
\citep{Dou16}.  The purposes of the light curve fitting are not only to measure 
the characteristic time scales of MIR variabilities, but also to determine the 
quiescent MIR emissions from the host galaxies when there are no pre-flare MIR 
photometries of the host galaxies. The exponential model gives a good fit to LCs 
for all objects except for SDSS J155223.29+323455.1, for which the highest observed 
flux data point may be still on the rising part of LCs due to the large gap between 
ALLWISE and NEOWISE-R.  

The decline time varies from 0.2 to 9 years with a median of 0.75 years, and the 
peak MIR luminosities of the flare after subtracting the host galaxy\footnote{We 
will always refer the flare luminosity to the the host galaxy subtracted MIR 
luminosity.} are in the range of $10^{42}$ to $10^{44}$ \ergs~ with 85\% sources 
between $10^{43}$ and $10^{44}$ \ergs. Moreover, the decay time does not 
significantly correlate with the peak luminosity (Figure \ref{Ltau}). With the 
Spearman rank correlation coefficient of $\rho=0.492$ for 14 points, the chance 
probability is  10\%.  The Logarithmic integrated energy in MIR for black 
body model during the observed falling part of LCs are in the range of 
49.2 to 52.4 with a median of 50.6 (ergs). They are only lower limits to the 
total energy of flares.

\subsection{Constraints on UV luminosity\label{sec:LUV}}

In the dust emission scenario, one can estimate the primary UV luminosity by considering 
that fact that dust grain is essential a bolometer at thermal equilibrium.  For an isotropic UV 
emission source of luminosity $L_{\lambda}$, the equilibrium can be written as (e.g., Laor \& 
Draine 1993)   
\begin{equation}
\frac{L_{UV}<Q^{abs}> }{4\pi R^2}=4 \sigma <Q^{em}(T_d)> T_d^4 
\end{equation}
where $Q^{em}(T_d)=4\pi\int_{0}^{\infty} Q_a(\lambda) B_\lambda(T_d) d\lambda/(\sigma T_d^4)$ 
is the average dust absorption coefficient for re-radiation light;  $<Q^{abs}>=\int_0^\infty 
J_\lambda Q_a(\lambda) d\lambda/\int_0^\infty J_{\lambda}d\lambda$ is the average absorption 
coefficient for the UV source. $<Q^{em}(T_d)>$ is a function of  dust temperature while $<Q^{abs}>$ 
depends on the shape of UV continuum source. Noting that both coefficients are a  function 
of grain size and composition. In the case of TDE, the observed SED in optical to UV can be  
well fitted by a black body model with temperature around a few $10^4$ K \citep{Gezari09, 
Holoien16a,Holoien16b,Brown17}. In the following, we will assume that the primary UV source is a  
black-body of temperature of $T_{bb}\sim 2\times 10^4$K, and  $<Q^{abs}>$ and $<Q^{em}(T_d)>$ 
are taken from \citet{Laor93}.  A low $T_{bb}$ will significantly increase $<Q^{abs}>$  for 
small grains, but it changes little if grain size is larger 0.1 micron.  

Once the distance to primary continuum ($R$) is known,  one can estimate the 
bolometric luminosity of the primary source. Due to extra-light traveling time, the IR flare 
lags the continuum one by order of light crossing time of dust emission region and also 
is broadened by a similar amount time although the exact value depends on the 
geometry of the dust distribution. As we will see in the next subsection, no optical flare 
is detected in the light curve of these sources so we cannot estimate the lag. However, with 
the MIR light curve, it is possible to get the width of flare approximately by the time scale of 
decaying from the peak. These luminosities are in the range from $10^{43}$ to $10^{47}$ \ergs 
depending on the dust composition and grain size \ref{fig:LUV}. It should be note that the UV 
luminosity is smaller than the infrared luminosity for silicate grains of size less than 0.1 
micron, thus these models are essentially ruled out.      
 
\subsection{Temperature and Mass of Dust}

We estimate the dust temperature and mass using the galaxy-light subtracted mid-infrared 
fluxes in $W1$ and $W2$. The galaxy light is estimated in the last section based on the light 
curve fitting. Since we do not know either the dust composition or the size distribution, we 
calculate the temperature and mass of warm dust at the maximum observed flux for black-body
model and silicate grains of sizes 0.01, 0.1 and 1 $\mu$m. Future infrared spectra will allow better 
constraints on the dust properties and dust masses. It should be noted that the mass of warm dust 
also evolves with time, and it is not necessarily to be the largest at the peak of IR luminosity. 
The reason for using the peak MIR photometries and colours is to minimize the photometric errors 
which tend to be bigger at later times.

We fit both black-body ($B(\lambda,T)$) and thermal dust models ($Q_{abs}(\lambda) B(\lambda, T)$ 
to the galaxy background subtracted $W1-W2$. We adopt $Q_{abs}(\lambda)\propto\lambda^\beta$ with a 
$\beta=-0.72,\,-0.81,\,-1.31$ in the wavelength range 1-10 $\mu$m for silicate of sizes $a=$0.01, 0.1 and 
1.0 $\mu$m, respectively (Drain \& Lee 1984). The derived dust temperatures are lower than the 
sublimation temperature for silicate (900-1200K) or graphite (1500-1900K, e.g., Lebreton et al. 2013). 
Since one may expect that dust at the sublimation radius responses first to the UV flare 
and emits most of IR light, the lower dust temperature suggests that the optical depth / 
the covering factor of dust there is smaller than those at larger radii or WISE missed the 
luminosity peak. The former case implies a cavity of dust around the black hole. Since 
there is a low-density pc-scale cavity inside the central nuclear disk in the Galactic 
centre (e.g., Etxaluze et al. 2011), it should not be surprised if similar cavities present 
in other quiescent galaxies. On the other hand, the inner edge of dust torus in Seyfert 
galaxies is usually found around the sublimation radius (e.g., Kishimoto et al. 2007). 
The dust mass is obtained from the infrared luminosity in $W2$ band 
\begin{equation}
M_d=\frac{a\rho L_\lambda}{3\pi Q_{abs}(\lambda)B(\lambda,T_d)}.
\end{equation}

We adopt $\rho=2.7$ g~cm$^{-3}$ for silicate grains. The final results are shown in Table \ref{table2}. 
The mass of warm dust is in the range of 0.03-3 $M_\odot$, which is similar to the TDE candidates 
with ECLs \citep{Dou16}. This amount of dust is one order of magnitude less than that within 1 parsec 
central cavity of the Galactic centre, and 2 orders of magnitude smaller than in the inner central nuclear 
disc  \citep{Etxaluze11}. On the other hand, if dust is embedded in individual gas clumps, which is 
optically thick to UV and optical radiation, then only grains at the skin are heated. Furthermore, only 
part of dust is heated at a specific time if MIR is caused by dust echo of a short UV flare. The total 
amount of dust would be considerably under-estimated using our method. 

\subsection{Searching for Optical Flares\label{sec:crts}}

For objects in the list, we extract the light-curves from the second public data 
release (CSDR2) from Catalina Real-time Transient Survey (CRTS, \citet{Drake09} ). 
The light-curves cover the period between April/2005 and October/2013. We calculate 
a quarter median magnitudes for the light curve to increase the signal to noise ratio for 
each data point. Unfortunately, we do not find any significant variations in the optical light 
curve. The standard deviations of these quarter medians is from 0.01 to 0.06 magnitude, 
which are consistent with measurement errors. We can set an upper limit on the peak of 
optical flare during CRTS period using the 3$\sigma$ standard deviations in these quarter 
medians. We found that these upper limits ($\nu L_\nu^V$) are 0.2 to 1.2 dex smaller than 
the black body infrared luminosity in Table 2.

There are several possibilities for the lack of optical flare in the CRTS light curves:  

\begin{enumerate} 
\item Stellar light outshines the optical flare in the low spatial resolution optical image. One 
important question is that the non-detection is expected or not if MIR flares are due to TDE 
echoes.  In SDSS J0952+2143, \citet{Palaversa16} estimated the absolute peak magnitude 
of the flare in $r$ band is no fainter than $M_r\simeq -20$ mag in SDSS J0952+2143, while 
\citet{Wang11} assessed a lower limit to optical magnitude of flare $Mg<-17.3$ in SDSS 
J0748+4712. These converts into an MIR to optical luminosity ratio -0.1 and -0.8 dex, which 
seems in tension with above upper limits, considering that fact none has been detected. Similarly, 
Jiang et al (2017) found that the peak MIR luminosity is a factor of ten smaller than the peak 
optical luminosity in PS16dtm.  Only in the TDE candidate in ULIRG F01004-2237, the peak 
luminosity of flare in MIR is  about 0.8 dex higher than in optical, which is believed to be partly 
obscured (Dou et al. 2017). 

\item Optical flares occurred before CRTS started, i.e., before 2005. Noting this is conceivable 
considering that  in ECLE SDSS J0748+4712 and J0952+2143, optical flares taken placed 
in 2004 \citep{Wang11,Palaversa16}, at the same time WISE detected the infrared echo 
\citep{Dou16}.  If this is true, the size of infrared region is much larger than we estimated 
using the decline time scale, so does the estimate UV luminosity in \S \ref{sec:LUV}.  In two objects,  
we saw a fast rising phase in less than a year, that would be very difficult to be explained 
in this scenario as it requires a special geometry of dust distribution. Much dust distributes 
over a large range of the outside of a paraboloid with a time delay of order a few years, but 
little inside.   

\item Optical flares are obscured by dust in the galactic nuclei. Considering presence of warm 
dust emission, it is plausible that dust may block the line of sight in some objects. From statistics 
point of view, one must see optical flares in some sources if dust covering factor is not close to unity.

\item Optical flares may be considerably weaker than in ECLEs because of large black hole masses. 
It is interesting to note that the black hole mass distribution of our sample is significantly different 
from the optical selected TDE. In optical selected TDEs, the optical emission is thought to be produced 
through reprocessing at large radius \citep{Strubbe09,Piran15,Stone16}, e.g., thick super-Eddington disk wind. 
For some reasons, if such a reprocessor is absent for large black hole masses due to reduced Eddington 
ratio, optical emission should be weaker in these objects in comparison with their UV luminosities.

\item Lastly,  IR emission is caused by non-thermal process, instead of reprocessed dust emission, 
which is peaked in IR.  Although we limited our sample to radio weak sources in quiescent state, there 
is no simultaneous radio observation that can limit non-thermal emission during the flare. Certainly, future 
observations can constrain this. 
\end{enumerate}  

\section{The Host Galaxy Properties}

The SDSS images of these galaxies are displayed on the Figure \ref{images}. They are resolved with a 
Petrosian radius $2\farcs88<R_{Petro} <7\farcs5$.  According to the automatic morphology classification 
by \citet{Kuminski16}, which agrees with Galaxy Zoo debiased "superclean" data-set at 98\% probability, 
7 galaxies are more likely elliptical ($\geq 0.6$); 4 are spirals, and 3 are ambiguous (either probability is 
less than 0.7), and remained one is not included in their catalogues. So, ellipticals outnumber spirals. On 
the other hand, galaxies can be classified as disc-like or elliptical-like based on the probability of the 
fits to the 1-D profile by either exponential law or de Vaucouleurs law and on the concentration index 
$C=R_{90}/R_{50}$  (e.g.,Bernardi et al. 2003). Using the probability derived with SDSS pipeline, 
exponential model is preferred over the de Vaucouleurs-law for 3 galaxies in both $r$ and $i$-bands; 
for another only in the $r$-band. For the remaining objects, de Vaucouleurs-law gives a much better 
fit than the exponential law in both $r$ and $i$ bands. On the other hand, 5 galaxies do not meet the 
requirement of $C>2.5$ for an early type galaxy. Among them, 3 objects can be better fitted by an 
exponential law. Thus, we conclude that more than half of the sample are early-type galaxies, 
dominated by spheroid components. In the later analysis, we define 7 galaxies that satisfy all above 
criteria as early-type galaxies.  

Colour magnitude diagram (CMD) has been used to demonstrate the bimodal distribution of galaxies, which 
has significant implication for the galaxy evolution \citep{Bell04}. On CMD, galaxies are clustered into red 
sequence or blue cloud, and a few galaxies scattered in the green valley in between. Figure \ref{msigma} 
shows our galaxies on the CMD with low redshift ($z<0.1$) SDSS galaxies in the legacy survey in contours. 
We have applied k-correction (v3\_2) using the code given by \citet{Blanton03}, so both colour and magnitude are 
referred to the frame at $z=0$. We do not make $V/V_{max}$ correction, thus the density of low luminosity 
galaxies is under-estimated by a large fraction, as a result bimodality is much less obvious. We over-plot our 
galaxies on the same figure, and most our galaxies are located on the red sequence of the massive galaxies. 
It is interesting to mention that all non-early-type galaxies except one also show red colours. This is 
consistent with the spheroid-dominated galaxy morphology. The high luminosity of hosts indicates that 
these galaxies possess large SMBHs. 

\citet{Simard11} de-composed the 1-D profiles of 660,000 SDSS galaxies into disc and bulge components in 
$g$ and $r$-bands. \citet{Mendel14} extended the work to the rest SDSS-bands, and calculated the total, bulge 
and disc stellar masses. All galaxies in our sample are included in \citet{Simard11}, and bulge masses are available for 
all galaxies but SDSS J095858.5+021459 in \citet{Mendel14}. The total stellar masses of these galaxies are in the 
range $9.7\leq \log M_* \leq 11.5$ $M_{\sun}$, and the bulge mass consists of 30\% to 100\% total stellar mass 
of the galaxy.  In the sample, the S\'ersic index of the bulge ($n_b$) is less than 2 for two galaxies 
(100933.14+232255.8;123715.10+601207.0),  between 2 and 3 for another galaxy, and greater than 3 for the 
remaining 12 galaxies.  According to \citet{Fisher08}, bulges with $n_b<2$ are most likely pseudo-bulges, 
and those with $n_b>2$ are classical bulges. Therefore, most of these galaxies have classical rather 
pseudo-bulges. 

As a further check, we plot the stellar velocity dispersion versus the absolute magnitude in $i$ band 
(known as Faber-Jackson relation) in Figure \ref{msigma}. For comparison, we show the distribution 
of low redshift ($z<0.1$), early type galaxies \citep{Bernardi03} in SDSS legacy sample as contours 
on the sample plot. The stellar velocity dispersions ($\sigma_*$) are retrieved from the Portmouth 
emission line database \citep{Maraston13, Thomas13} in the SDSS archive. First, it should be pointed 
out that three galaxies in this sample has $\sigma_*<70$ km~s$^{-1}$, i.e., smaller than the spectral 
resolution of the SDSS. Their $\sigma_*$ may be not reliable, and one may take seriously its exact value, 
but rather treat them as $\sigma_*<70$km~s$^{-1}$. This does not affect our main results.  Second, 
these may not represent the $\sigma_*$ of the bulge for disc galaxies because the light within 3 arcsecs 
of SDSS fibre aperture may include significant contribution of the disc component. The effects depend 
on the angular size of the bulge and disc, the inclination of the disc, etc. Thus, there is no simple way to 
correct them. Nevertheless, considering the fact that the bulge component accounts for a substantial 
fraction of total light even for the object with the least bulge to disc ratio, and dominate in more than 
half of the objects. Thus, we will use them as an approximation for the $\sigma_*$. For early type 
galaxies, there is an additional effect that $\sigma_*$ is usually defined as being measured with either  
$r_{eff}$ or $r_{eff}/8$. However, considering the weak dependence on $\sigma_*$ on aperture size 
$\sigma_*\propto (r/r_{eff})^{-0.08}$ when $r$ is small (e.g., Falc\'on-Barroso et al. 2017), the aperture 
corrections have only small effect on the Faber-Jackson relation considering its relative large intrinsic 
scatter. In comparison with SDSS early-type galaxies, 7 early-type galaxies in this sample 
appear to have smaller stellar velocity dispersions at a given optical magnitude. The offsets in 
Faber-Jackson relation is in line with the offsets in the BH masses derived using above two 
methods. 

The deviation from the Faber-Jackson relation can be due to either an excess of galaxy luminosity or a 
deficiency in stellar velocity dispersions. In the first case, objects with too high luminosity may be due to 
younger stellar population or low metallicity. To check this, we plot the $b-r$ vs $M_i$,  and compare it to 
the low-redshift ($z<0.1$) early-type galaxies in Figure \ref{msigma}. Apparently, early-type galaxies in 
our sample have similar colours as others, suggesting that stellar age or metallicity is not the main cause for 
such deviation. The over-luminosity may also be caused by many dry minor-mergers during the accumulation 
of extended stellar envelope, while the compact core remains almost unshaped \citep{Huang13}. As accreted 
small galaxies are usually bluer and have a lower metallicity, the add-on stellar component is expected to be 
bluer \citep{Huang16}, which does not agree with above analysis. Thus, this seems also not likely the main 
reason. 

Alternatively, the stellar velocity is under-estimated for the mass, probably caused by different evolution track. 
Smaller stellar velocity dispersions suggest that these systems have relatively higher energy than average early 
type galaxies according to virial theory. Simulations showed that during dry-merging of two galaxies, the orbit 
energy of the system transforms into the internal energy, leading to a puffed-up remnant, especially on nearly 
radial orbits \citep{Boylan06}. During this process, the colour of galaxy should not be affected, in accordance with 
the above analysis.
 
 We show the spectral energy distributions (SEDs) of these galaxies in Figure 
 \ref{fig:sed}.  The  SED is peaked  0.7- 2 $\mu$m and then decreases steeply 
 to the 4-5 $\mu$m in the low state, in particularly, suggesting the dominant old  
 stellar populations.  About 2/3 display another peak in 10-20 $\mu$m or longer, 
 implying a moderate star-formation in these galaxies.  

\section{Estimate of Black Hole Masses}

It is well established that masses of SMBHs ($M_{BH}$) are tightly correlated with 
the stellar velocity dispersions ($\sigma_*$), the masses ($M_{bulge}$) and the K-band 
luminosities ($L_K$) of the classical galaxy bulges (see Kormendy \& Ho 2013 for a review, 
hereafter KH13).  However, it is still controversial whether black hole mass is 
more tightly correlated with $\sigma_*$ or equally well with $M_{bulge}$ and $L_K$ (Graham 
2016 for review), or whether different type of galaxies (e.g., core or power-law galaxies) follows the same 
correlations (e.g., McConnell \& Ma 2013, MM13 hereafter; also KH13 and reference 
therein). In either KH13 and MM13, the bulge mass is calibrated using a mixture of stellar 
and dynamic mass, which is not available for our objects. So  we will use $M_{BH} -\sigma_*$ 
relation as the base black hole mass estimate.
 
In the following, we will adopt $M_{BH}-\sigma_*$ relation in MM13. The slope 
in MM13 ($\beta=5.20$) is steeper than those in KH13 ($\beta=4.26$), and the two relations 
agree at $\sigma_*=235$ km~s$^{-1}$.  For the infrared flare sample in this paper, 
the black hole derived using MM13 is 0.8 dex smaller than those using KH13 for 
the smallest $\sigma_*$, while the difference is less than 0.5 dex for majority. We 
use MM13 because it is more consistent with recent work of van den Bosch (2016). 
With this method, the BH masses are in the range from below $1.4\times 10^4$ 
to $4\times 10^8$ $M_{\sun}$ with a median of $2.5\times10^7$  $M_{\sun}$. It 
should be noted that black hole masses less than $10^6$~$M_{\sun}$ are not reliable 
because their $\sigma_*$ is below the SDSS spectral resolution, so they can be taken 
as upper limits of $10^6$~$M_{\sun}$.   

We find that the observed maximum luminosity of infrared bursts is fairly well 
correlated with the black hole mass from $\sigma_*$ (Figure \ref{lir_host}). 
A Spearman rank correlation analysis gives a correlation coefficient 
$\rho=0.758$,  corresponding to a null probability of $2\times 10^{-3}$. 
A linear fit yields the following relation $\log L_{IR} ({erg~s}^{-1})=(0.33\pm 0.07)\times \log 
(M_{BH}/10^7~M_{\sun})+(43.36\pm 0.08)$. The flat slope means that MIR Eddington 
ratio decreases with black hole masses.  The MIR luminosity is also correlated with 
the stellar mass of galactic bulge  ($\rho=0.789$ and $P_{Null}=8\times 10^{-4}$) 
and the absolute optical magnitude of the galaxy at $i$ band ( $\rho=-0.710$ and 
$P_{Null}=0.004$, Figure \ref{lir_host}).

\section[]{Estimate of Event Rate} 

The accumulative number increases with redshift roughly proportional to the 
comoving volume within the SDSS footprint up to redshift $z=0.12$, then 
flattens (Figure \ref{redshift}) suggesting that the sample becomes seriously 
under-representative beyond that redshift assuming that there is no abrupt 
evolution in the event rate. So in the following, we will only consider the subsample 
of 10 objects at redshift $z\leq 0.12$. This gives a density of about $8\times 10^{-8}$ 
Mpc$^{-3}$. However, as we have already mentioned, requiring $var\_flag\geq 7$ 
selected only one of 4 TDE candidates with ECLs, which all showed long term declined 
light curves with ~$\Delta W2>0.3$ mag in the ALLWISE data. So, the density is likely a 
factor 4 higher.  
   
To obtain the rate, we need to know the average time-scale that the flare can be 
picked up with our criteria. This relies on our understanding of various selection 
effects imposed by our criteria, the WISE sensitivity and observation cadence 
etc, as well as the parameters of light curves, such as the statistical distribution 
of the peak luminosity and decay-time scales. Because these properties are unknown, 
here we will only give an order of magnitude estimate. By noting that all objects in our 
sample display a ~$\Delta W2> 0.2$ mag during the 3-years ALLWISE period, we take the   
duration of the light-curve that W1 or W2 magnitude changes at least 0.2 mag within 
3 years as the appropriate time scale. For most objects, we only observed the declining 
part of the light curve, and thus can only give a lower limit to time scale. The average 
value is about 3 years. Putting these together, we attain an event rate of order 
$10^{-7}$ Mpc$^{-3}$~yr$^{-1}$.

However, we showed in last section that the event does not take place at the same 
probability for each galaxy. Rather, host galaxies mostly are red and luminous. 
Guiding by eyes, we draw a straight line on the CMD-diagram (Figure \ref{msigma}). 
All but one locate on the upper right of the line. Now we count all spectroscopic 
galaxies on the same regime of CMD-diagram in the SDSS DR-12 spectroscopic catalogue 
in the redshift $z<0.12$. The number of galaxies turns out to be 62710. Taking into 
account the duration of the light curve that can be detected and the fraction of 
sources missed due to variability selection criterion discussed above, we estimate 
the rate of event around $2\times10^{-4}$ gal$^{-1}$~yr$^{-1}$ for red-luminous galaxies 
defined in the Figure. We will leave a more thoroughly statistical study of the event 
rate to a later work using a sample of galaxies without such restrictions.    

\section[]{Discussion}

We have identified 14 normal galaxies which display flare-like mid-infrared 
light curves by combining the SDSS spectroscopic data and the WISE multi-epoch 
photometries. By selection, these sources are required to have normal galaxy 
optical spectra, with either star forming emission lines, early type 
absorption features or LINER.  Because W1 and W2 show the same trend, they 
are not  spurious. Mid-infrared variabilities could have different physical sources, 
such as supernova explosion, or fluctuation of accretion disc in an obscured 
Seyfert galaxy, or sporadic accretion onto a supermassive black hole due to 
instability in an accretion disc or a tidal disruption of a star. 
We will discuss these possibilities in detail. Any likely scenario should be 
able to explain the following observations: (1) lack of Seyfert-like narrow line 
spectrum; (2) much higher probability of finding flare-like light-curves among 
MIR variable non-Seyfert galaxies than among MIR variable Seyfert 
galaxies; (3) the maximum infrared luminosity in the range of $10^{42-44}$ \ergs 
and the integrated MIR flare energy in the range of $10^{49-52}$ ergs; (4) 
typical decay time is a year, and rising time is an order of magnitude shorter; 
(5) host galaxies are mostly red and luminous, and black hole masses distribute 
in a rather broad range with more than half in the range of $10^7$ to $10^8$~
$M_{\sun}$ and one larger than $10^8$~$M_{\sun}$; (6) the maximum infrared 
luminosity is correlated with the black hole mass or the stellar mass of galactic 
bulge, and typical MIR Eddington ratio is between $10^{-2}$ to $10^{-3}$; 
(7) there had been no detectable optical flare in the CRST data from 5 
years before the appearance of infrared flare, with the limits on the luminosities  
of optical flares to 0.2 to 1.2 dex smaller than of these MIR flares. 
If the infrared light comes from thermal emission of dust heated by UV light 
from an accretion disc around black holes and assuming a dust covering factor 
of 0.1, about 0.002-1.4 $M_{\sun}$ gas had been accreted during the burst 
assuming a typical radiation efficiency of 0.1. 

\subsection{Non-Thermal Flares from Low Power Blazars}

Could it be possible that the infrared flares are associated with blazar-like activities 
that below the radio detection limit in quiescent state? In the case of non-thermal 
emission, it is more proper to use the spectral index rather than black-body temperature 
to characterize the spectrum. The flux ratios of W1 to W2 bands during the peak give 
power-law indices $\alpha$ ($f_\nu\propto \nu^{-\alpha}$) in the range of  0 to 3.2, with 
a median of 1.8, which are within the range of blazars \citep{Giommi94}.  9 of 14 
sources have $\alpha>1.6$, they should be low synchrotron-peaked ones if they are 
blazars.  As we discussed in \S \ref{sec:crts}, an analysis of CRTS light curves set upper 
limits on the luminosity of optical flares to be 0.2 to 1.2dex lower than these of MIR flares. 
This is consistent with SEDs of low- to intermediate-synchrotron-peaked blazars, but is 
inconsistent with those high peak blazars unless the optical light is heavily extincted by 
dust \citep{Rani11}. However,  there are several problems for blazar-like non-thermal 
activity. First, radio weak blazars are usually high synchrotron-peaked blazars rather 
than low synchrotron-peaked ones \citep{Giommi94,Mao16}. Second, variability flag 
selected blazars show much lower chance (9.2\% versus 61\%) of flare-like MIR light 
curves than the non-Seyfert galaxy sample. Thus, we consider that blazar-like activity 
is less likely.

\subsection[]{Infrared Luminous Supernova} 

Infrared emission has been detected in about 20\% nearby sample of 
supernovae (SNe) at a few months to several years after the explosion 
using Spitzer IRAC \citep{Fox11, Fox13, Tinyanont16}. The peak infrared 
luminosity lies in the range of $10^6$ to $10^9$ $L_\odot$, depending on 
the type of SNe. Type Ia SNe usually shows very weak mid-infrared emission 
and is not detectable three years after the explosion\citep{Tinyanont16}.
type IIn SNe are more luminous, and the infrared emission may last up to ten 
years due to the heating of pre-existing dust by radiative shock between  
the expanding SN shell and the dense wind of the progenitor \citep{Fox13}. 
We have cross-correlated with the SN catalogue \citep{Guillochon16}, 
and does not find any matches within 10 arcsecs matching radius. This 
appears not surprising as it is difficult to find SNe in galactic nuclei. 

Although the light curves of our objects look similar to those of SNe,  
SN scenario falls in two aspects.  First, the peak MIR luminosities are much 
higher than those of SNe (Figure \ref{lumdist}). Among 14 objects, all but one 
have MIR luminosities greater than $3\times 10^9 L_\odot$, while among 22 
type IIn SN detected by Spitzer and WISE, only one is above $10^9 L_\odot$ 
(Fox et al. 2013). It is even remarkable considering the fact that most of our galaxies have 
red colours, so current star-formation rates are low, thus most SNe in 
these galaxies should be type Ia. The long lasting time of infrared 
burst for more than five years is not compatible with known type Ia 
supernova. Second, it is very difficult to understand the strong correlation 
between the peak luminosity and the mass of SMBH in the centre of the 
host galaxy in the scenario of supernova. 

\subsection{Fully Obscured Seyfert Galaxies}

Is it possible that those 14 galaxies in our sample are heavily obscured 
Seyfert galaxies?  In this scenario, the observed MIR LCs could be naively 
due to AGN activities in Seyfert galaxies. However, this scenario cannot 
explain two facts. First, there are no optical signatures suggesting 
existence of Seyfert nuclei,  AGN-like narrow line spectra.  Second, the 
MIR LC morphology and the time scales observed among our sources are 
quite different from stochastic behaviours of AGN LCs. Finally, 
we have rejected objects with infrared colours of typical Seyfert galaxies.

Lack of optical spectroscopic signatures of Seyfert galaxies can be caused 
by galactic scale obscuration. In the classic picture of type-2 AGN, the broad 
line and continuum regions are obscured by dust tori, while the extended 
narrow line regions, so called ionization cone, are less obscured and more 
visible. However, if there is a large-scale dust lane in the galaxy, the narrow 
line region may be also obscured, thus optical AGN signatures based on narrow 
line ratios could become less clear due to dust extinction. With less dust obscuration 
in the mid-IR, we may still be able to see mid-infrared emission from inner regions. 
Mid-infrared variability thus is expected. We believe that WISE variable normal 
galaxies with significantly fluctuating light curves are buried Seyfert galaxies. As 
we have already showed that about 15\% type 1 AGNs display flare-like light-curves, 
there should be no surprising that some of our objects are also obscured Seyfert 
galaxies. Assuming the obscured and non-obscured Seyfert galaxies has the same 
probability of showing flare-like light curve, we would expect 2-3 fully obscured AGNs 
among of our sample. They would show strong mid-IR excess, a type-1 AGN feature 
in the infrared, and probably some of them have already been removed using the 
WISE colour ($W1-W2\gtrsim 0.75$).
  
\subsection{Outburst of Low Accretion Rate AGNs}

In this scenario, the central black hole is episodically fed at a high 
accretion rate for a relatively short time scale probably caused by 
instability in the accretion flow and then return to the quiescent level. 
In this case, the narrow line region may be too small to have significant 
emission in [O {\sc iii}] \footnote{\citet{Yang13} detect brightening of [O {\sc iii}] in the 
TDE candidates with ECLs} to make a Seyfert-like spectrum. However, low 
level activity in the quiescent state may also explain their very common 
LINER spectra. Regardless of the physical process responsible for such 
episodic fuelling, the rising and decay time must be shorter than the 
viscous time scale of the disc at the radii causing dramatic change of the 
accretion rate. For geometrically thin accretion disc, the viscous time 
scale can be written as \citep{Czerny06} 
\begin{equation}
t_{visc}=0.32\alpha_{0.1}^{-1}(r/10h_d)^2r_3^{3/2}M_8 \mathrm{(yr)} 
\end{equation} 
where $\alpha_{0.1}$ is dimensional viscosity coefficient,  $h_d$ the disc height,  
$r_3=r/3R_{sch}$ and $M_8=M_{BH}/10^8M_\odot $. In the radiation dominated 
inner disc, $h_d \simeq 10\dot{m} R_{Sch}$. Thus, the short observed 
rising and decay time (years) suggests that the perturbation must occur at a 
small radii (order of $10 r_{Sch}$ for a few $10^7$ solar mass black hole and 
Eddington ratio of 0.1). Yet the duty cycle of the event must be very low in 
order not to produce Seyfert-like narrow emission lines. From the rate estimated 
in last section, such event occurred once in $10^4$ yr in luminous red galaxies, 
consistent with this requirement.

In black hole X-ray Nova, the X-ray flux rises in several days by a 
factor of several hundreds and last for several tens to hundred days 
before declining to the pre-burst level (Tanaka \& Shibazak 1996 for a 
review). It is general accepted that the thermal instability due to 
hydrogen ionization in the thin accretion disc is thought to trigger a 
transition from a cold, low accretion-rate branch to a warm, high 
accretion-rate mode (See Lasota 2001 for a review). The mechanism operates 
when the steady income accretion rate is lower than that is required to 
sustain a steady high accretion mode. When scaling to the black hole mass 
here, the rising time (thermal time scale at a fixed disc temperature 
$\tau_{rise}\propto M^{1/3}$) would be one to a few years for black hole 
mass of $10^7$ to $10^8$ $M_{\sun}$ if the dimensionless quiescent accretion 
rate is similar to those in X-ray Nova \citep{Lin86}. 
If the quiescent accretion rate is as low as $10^{-4}$, the ionization 
instability radius may extend to 13 $r_g$ for a $10^8$ $M_{\sun}$ black 
hole \citep{Janiuk11}.  As we estimated at the begin of this section, 
the total mass accreted during the outburst is order of one solar mass from 
observation. This is in coincidence with the disc mass in a wide annulus of 
$r=10\;r_g$ at the critical accretion rate (See Figure 1a of \citet{Lin86}  for
 the disc column density). However, their model predicts only a moderate 
 increase in the accretion rate with a duration of the burst (tens to hundred 
 years) much longer than observed. In addition, it is also questionable whether 
 the accretion flow can keep geometrically thin down to such small radius at 
 such low accretion rate \citep{Yuan14}. Despite all of theoretical uncertainties, 
observations show that some quasars and Seyfert galaxies turn on/off on time scales 
of years \citep{Gezari16}, suggesting that such episodic accretion does occur on 
time scales of years in AGNs, but the observed frequency of these events is low. 

\subsection[]{Echoes of Tidal Disruption Flares}

Giving the recent detection of the reprocessed TDE infrared emission 
\citep{Jiang16, vanVelzen16}, it is likely that some or most of  the MIR long 
fading galaxies are of the same nature. The observed MIR 
luminosity is about a few times $10^{41}$ \ergs\ for ASSASN-14li 
\citep{Jiang16} and a few $10^{43}$ \ergs\ for the TDE candidates with 
ECLs \citep{Dou16}. The observed peaked infrared 
luminosities of our objects are in the range of $(0.2-10) 10^{43}$ 
\ergs\ with a median of $3\times 10^{43}$ \ergs\ using the black-body 
fit to {\it W1} and {\it W2} bands, which is similar to the TDE 
candidates with ECLs. 

As already noticed in \citet{vanVelzen16}, about 60\% bright 
TDEs were detected in reprocessed infrared emission by WISE.  Had not 
ASSASN-14li been so close, it would not have been detected by WISE 
\citep{Jiang16}. On the other hand, in gas rich environment, such 
as ECLEs, typical MIR luminosity is a few $10^{43}$ \ergs\, and the 
detection rate of 100\%. The high luminosity of these sources suggests 
that they reside in more gas-rich environment although we have the optical 
light-curve of only one source in the interesting time range (from SDSS 
spectroscopic observation to the WISE observations). \citet{Palaversa16} 
found that the ECLE SDSS J0952+2143 is a factor of two brighter in luminosity 
than PS1-10jh, led them speculating that it could be a more energetic event. This 
may also account partly the high infrared luminosity.  
  
However, we notice several differences in properties of host galaxies our objects 
and those of TDE candidates with ECLs or known TDE candidates. We collect the 
optical magnitudes of host galaxies for known TDEs or TDE candidates, and shown 
them also on the CMD. It is apparent that our sample has a distinct distribution than 
known TDEs or subsample of TDE candidates with ECLs. The host galaxies of our 
sources are more luminous and redder than the hosts of known TDEs (see Figure 
\ref{msigma}), while TDE candidates with ECLs are similar to the other known TDEs. 
For example, 7 of 14 galaxies in our sample are brighter than $M_i<-21.75$ mag, 
while only 3 of 39 known TDEs/TDE candidates do. Furthermore, those three galaxies 
in latter group all show blue colours ($g-r<0.6$), but 7 luminous galaxies in our sample 
all possess red colours ($g-r>0.65$). The red colour is unlikely due to the dust 
reddening to the stellar light but rather attributed to the old stellar populations because 
the reddest galaxies in the sample shows prominent 4000\AA~break. In line with this, 
these galaxies do not offset from the CMD of other SDSS galaxies. 

The red colour and high luminosity suggest that these galaxies host larger SMBHs 
than known TDEs or TDE candidates. In section 4, we derived $M_{BH}$ from 
$10^5$ to $4\times10^8$ $M_{\sun}$ using $\sigma_*$.  These masses are 
compared to a complete sample of 12 optically/UV selected TDE host galaxies
measured by \citet{Wevers17} in Figure \ref{fig:mbhdist}.  Apparently, the distribution 
of our sample is much more spread with much more objects in either larger or small 
masses. 10 of 12 objects in their sample have masses between $3\times 10^5$ and 
$10^7$ $M_\odot$, while only 2 of 14 objects in our sample fall in this range and most 
of them are from $10^7$ to $10^8$ $M_\odot$. 
Note that also one object has a black hole mass larger than $10^8$  $M_{\sun}$, 
which is the upper limit for the tidal disruption of a solar-mass main-sequence star around a 
Schwarzschild black hole. In the TDE scenario, this can be disrupting a solar-mass 
main sequence star by a spinning black hole \citep{Kesden12,Leloudas16, Margutti17} 
or disruption of a post main sequence star \citep{Kochanek16}. The different mass 
distribution from optical/UV selected TDE suggests that our sources are either not TDE 
or a population of yet missed TDEs. In passing, we note that some theoretical models 
of \citet{Kochanek16} predicted that the observed TDEs should be dominated by black 
hole masses between $10^7$ to $10^8$ $M_\odot$ although the true distribution is very 
uncertain. 
 
There are several potential drawbacks for the TDE scenario. As we discussed in 
the last section, the small stellar velocity dispersion at a given galaxy luminosity implies 
a less compact structure following fundamental plane \cite{Bernardi03}, which also 
disfavours the tidal 
disruption scenario. But if the excessive luminosity is due to accumulation of 
outer part of disc through the accretion or minor merger as discussed in last section, 
then this has little effect on the structure of galactic centre, hence the rate of TDE. 
Second, the red colour of the host seems at odds with the conventional wisdom that 
there is a dearth of cold gas and dust in red galaxies, although \citet{Young11} 
detected CO emission in at least 22\% early type of galaxies in the volume-limited 
ATLAS$^{3D}$ sample. 

On the other hand, there are some advantages for detecting infrared TDE 
echoes in early type galaxies. First, early type galaxies show systematically 
weaker mid-infrared emission than late-type galaxies. Thus it may be easy to 
detect infrared echoes. Second, these galaxies also host a more massive 
central black hole, thus a more luminous TDE is expected. For the disruption 
of a solar type star, the fallback rate decreases as the black hole mass increases 
($\propto M^{-1/2}$) in the Newtonian treatment. At black hole masses less than 
$3\times 10^7$ $M_{\sun}$, the rate of initial fallback exceeds the Eddington rate 
\citep{Rees88}.  Strong outflows are expected to be launched, and the radiative luminosity
may be still limited by the Eddington one \citep{Jiang14}, which increases with 
the black hole mass. Wevers et al. (2017) found that observed peak luminosity is 
consistent with Eddington limit accretion for black hole mass less than 
3$\times 10^7$ $M_\odot$. At black hole mass larger than this, the disruption
radius is so close to the black hole that relativistic effects must be taken
into account and the mass falling rate can be doubled \citep{Kesden12}, so the
peak bolometric luminosity may depend on the black hole mass very weakly. If
the infrared luminosity is proportional to that of bolometric luminosity, one
would expect that infrared luminosity increases with the black hole mass, and
then flattens. 

As we have mentioned that the non-detection of any optical flare in CRTS 
sets  a significant upper limit on the peak luminosity of the optical flare to be lower 
than expected for either SDSS J0952+2143 or PS16dtm with similar MIR luminosities 
(see \S \ref{sec:crts}).  This can be interpreted as either heavily obscured TDE or weaker 
than usual optical-flares. Note that optical emission in known-TDEs is stronger 
than the disc model predicted, and is assumed to form through reprocessing 
of the accretion disc radiation by an optically thick gas-shell with much larger 
radius than the disc (e.g., Stone et al. 2016). The reprocessor can be outflows 
launched by the super-Eddington accretion flows \citep{Strubbe09} or in the 
stream-stream collisions \citep{JiangY16}. We may 
expect that outflows weaken as mass infalling rate of debris becomes 
sub-Eddington when black hole mass increases to larger than $10^7$ $M_{\sun}$, 
so less hard disc radiation is reprocessed into optical light. 

With all pros and cons, it is inconclusive whether they are infrared
echoes of TDE. But it is clear that if these are indeed TDEs, they will
represent a subset that mostly missed by the current TDE surveys.  

\section[]{Summary}

We have examined 31 galaxies spectroscopically classified as star
forming galaxies, LINERs or absorption line galaxies that flagged
as variable in the ALLWISE catalogue. We find that
19 (61\%) of them show light-curves similar to IR echoes of tidal
disruption events, i.e., long term decline,  and in two cases also a 
relatively fast rising ahead. In comparison, among Seyfert galaxies
with variable MIR emission, the fraction showing such a light-curve
is much lower ($\sim 15$\%).  We carry out a comprehensive study of
these 19 sources using WISE archival data. Among them, 
5  have [W1-W2] colour $>$ 0.8, indicating them like being obscured 
AGNs. They were rejected from the sample. The maximum 
infrared luminosities after subtracting the host galaxy are in the ranges 
of a few time $10^{42}$ to $10^{44}$ erg~s$^{-1}$ with a median 
$4\times 10^{43}$\ergs, and correlate with black hole masses and the 
stellar mass and absolute magnitude of the host bulges. We fit the light 
curve with exponential law and the typical decay time is a year. We estimate 
warm-dust mass in the range of 10th to a few solar masses and temperatures 
from 600 to 1200 K at the maximum infrared flux.  With typical gas to dust ratio of
solar, we expect at least a few to a few hundred solar mass gas within
a distance of a few parsecs. The estimated dust mass is still significantly lower
than that observed in the inner parsecs of the Galactic centre.
Most of the host galaxies are intrinsically red and luminous, and
tend to have lower stellar velocity dispersions than galaxies with the 
same optical luminosity. We estimate a rate of event about $10^{-4}$
gal$^{-1}$~yr$^{-1}$ among red luminous galaxies. We do not detect any 
optical flares in CRTS light curves spanning 8.5 years from 5 years before the 
starting of the MIR survey, with upper limits on the flare luminosity in 
the V-band about 0.2 to 1.2 dex smaller than the MIR luminosities.

We considered several possibilities, including infrared echoes
from supernova, fully obscured Seyfert nuclei, the episodic
accretion due to certain instability in a low accretion rate
system, and echoes of TDEs. The strong correlation with the
host galaxy luminosity suggests that it is not supernova; also
the infrared luminosities are generally too high for SNe.
The scenario of fully obscured (including narrow line region) Seyfert
galaxies cannot explain the high fraction of flare-like light-curves
among infrared variable normal galaxies. While the ionization
instability of thin disc predicts much longer MIR flare even
at a very low accretion rate, and the physical mechanism of accretion
flow instability is still to be identified for episodic fueling for 
shorter time scales. The observed flare may be an analogy
to the turn on/off AGNs discovered in recent years but with even shorter 
time scales. In this picture, common LINER spectra are expected. 
The strong correlation between the peak luminosity and the black hole mass 
suggests that the dimensionless accretion rates at the peak are similar.
Although infrared luminosities and light curves are similar to that of known 
TDEs and TDE candidates with ECLs, their host galaxies are significantly 
more luminous and redder. If these events are indeed dust echoes of TDE 
optical flares,  they must be a subset of TDEs that are mostly missed by 
previous surveys. Black hole mass distribution has a peak between $10^7$ to 
$10^8$ $M_{\sun}$, significantly above that of TDEs selected in optical/UV.  
Non-detection of any optical flare in CRTS suggests that either optical 
flares are either intrinsically weaker than known TDEs with similar MIR 
luminosities or the nuclei are obscured.  

\section*{Acknowledgments}

We are grateful the referee for careful reading the manuscript and critical  
comments, that lead significant improvement of the presentation. We thank 
Dr N.C. Stone for helpful discussion. This project is supported by National
Basic Research Program of China (grant No. 2015CB857005), the NSFC through
NSFC-11233002, NSFC-11421303, and U1431229 and U1731104, jointly supported 
by Chinese Academy of Science and NSFC. This research makes use of data products 
from the {\it WISE}, which is a joint project of the University of California, Los Angeles and 
the Jet Propulsion Laboratory/California Institute of Technology, funded 
by the National Aeronautics and Space Administration. Funding for SDSS-III 
has been provided by the Alfred P. Sloan Foundation, the Participating 
Institutions, the National Science Foundation, and the U.S. Department of Energy Office of Science. The
SDSS-III web site is http://www.sdss3.org/. SDSS-III is managed by the
Astrophysical Research Consortium for the Participating Institutions of the
SDSS-III Collaboration.

\clearpage
\thispagestyle{empty}
\begin{figure*}
\centering
\begin{minipage}[c]{\textwidth}
  \centering
     \includegraphics[width=6.0in]{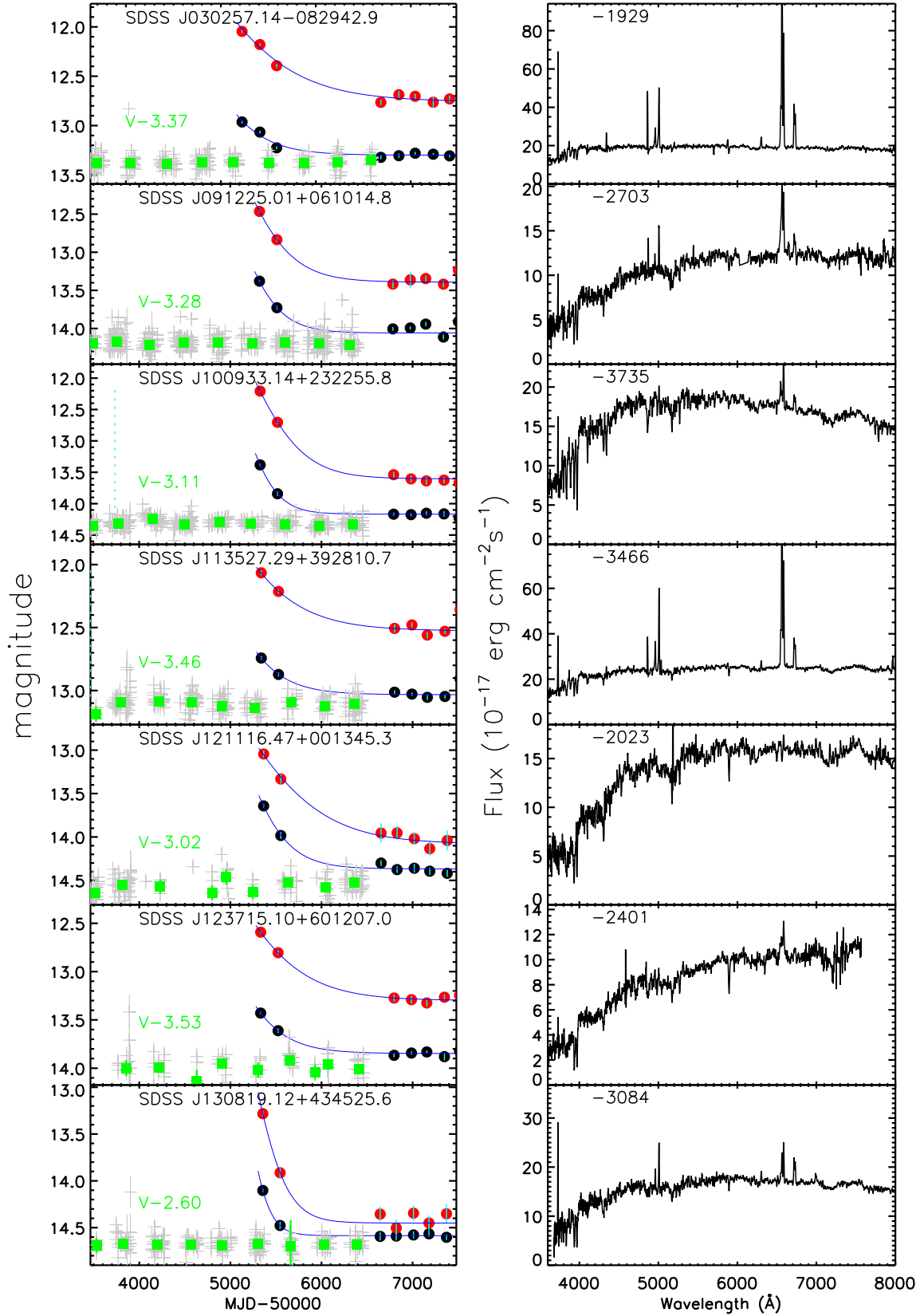}
     \caption{The SDSS spectra (right panels), and WISE and CRTS V-band light curves (left panels) 
     of sources in our sample. Left panels: the original CRTS light curves showed as grey bar, while 
     the quarter medians are displayed in green squares. The light curves in W1 and W2 are 
     represented with dark and red circles. The blue line shows the exponential-decay. 
Right panels: the number on the upper corner gives 
     the time gap between the first WISE observation and the latest SDSS spectroscopic observation; 
     a negative value means that spectrum was taken before WISE observation. When there is a second 
     spectrum of the object, we plot it in red. }
     \label{lcsp}
\end{minipage}
\end{figure*}
\setcounter{figure}{2}

\begin{figure*}
\centering
\begin{minipage}[c]{\textwidth}
  \centering
     \includegraphics[width=6.0in]{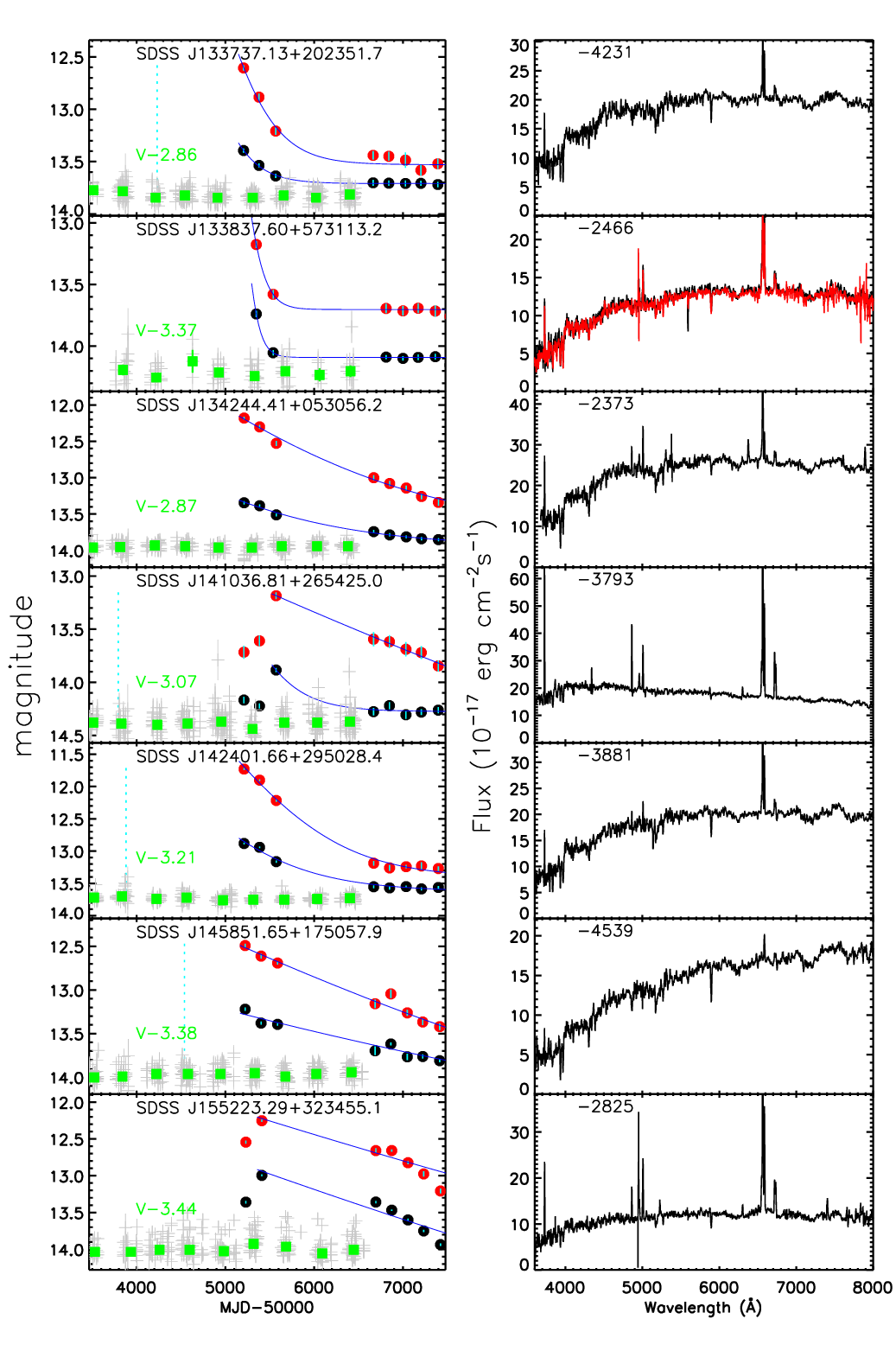}
     \caption{Continued.}
\end{minipage}
\end{figure*}
  
\thispagestyle{empty}
\begin{figure*}
\centering
\begin{minipage}[c]{\textwidth}
  \centering
     \includegraphics[width=6.0in]{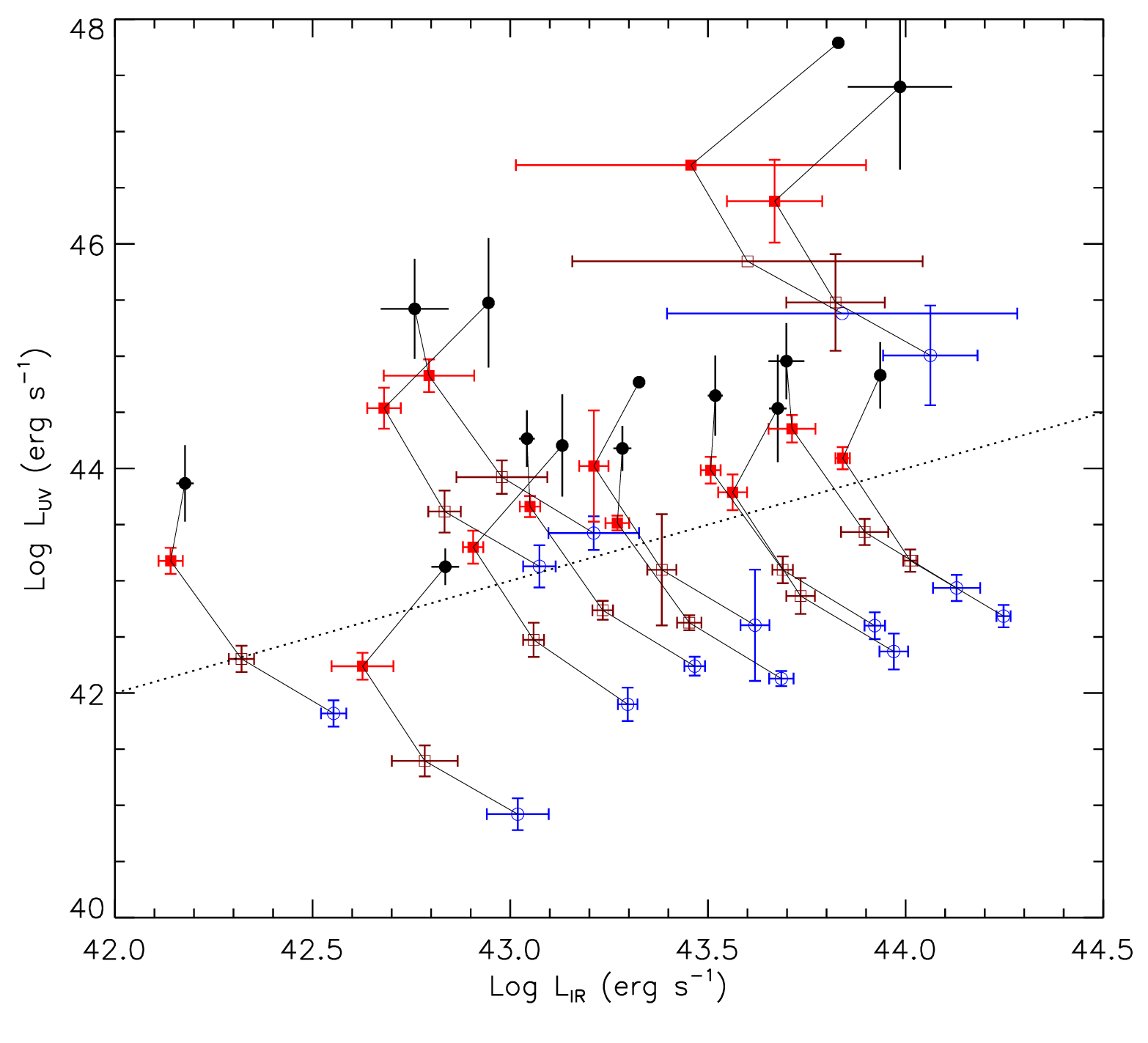}
     \caption{The primary UV luminosity versus observed IR luminosity for different black body 
     (filled black circles), silicate grains of size 1 (filled red squares), 0.1 (open brown squares) and 0.01$\mu$m (open blue circles). Different 
     models of the same object are connected by line. The dotted line in the figure represents the 
     equality of the IR and UV luminosity. Note that for an optically thick spheric dust shell of 
     radius R, the peaked IR luminosity is the average burst UV luminosity over a time interval 
     of 2R/c. }
     \label{fig:LUV}
\end{minipage}
\end{figure*}

\begin{figure*}
  \centering
  \begin{minipage}[c]{\textwidth}
  \centering
     \includegraphics[width=5.0in]{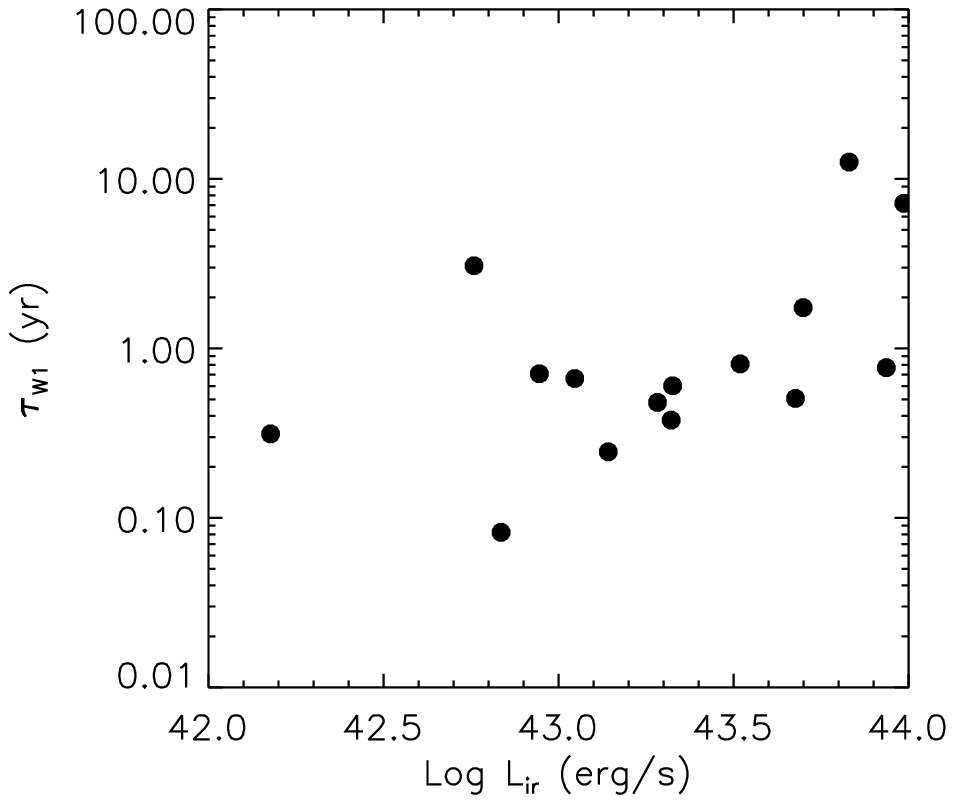}
     \caption{The correlation between the host galaxy subtracted MIR luminosity and the 
      decay time. The MIR luminosity is derived using the black-body model, 
 and the decay time is obtained in the exponential fit to the light-curves (see Table 2).  
      }
    \label{Ltau}
\end{minipage}
\end{figure*}

\begin{figure*}
  \centering
  \begin{minipage}[c]{\textwidth}
  \centering
     \includegraphics[width=7.0in,trim=0 5cm 0 0,clip]{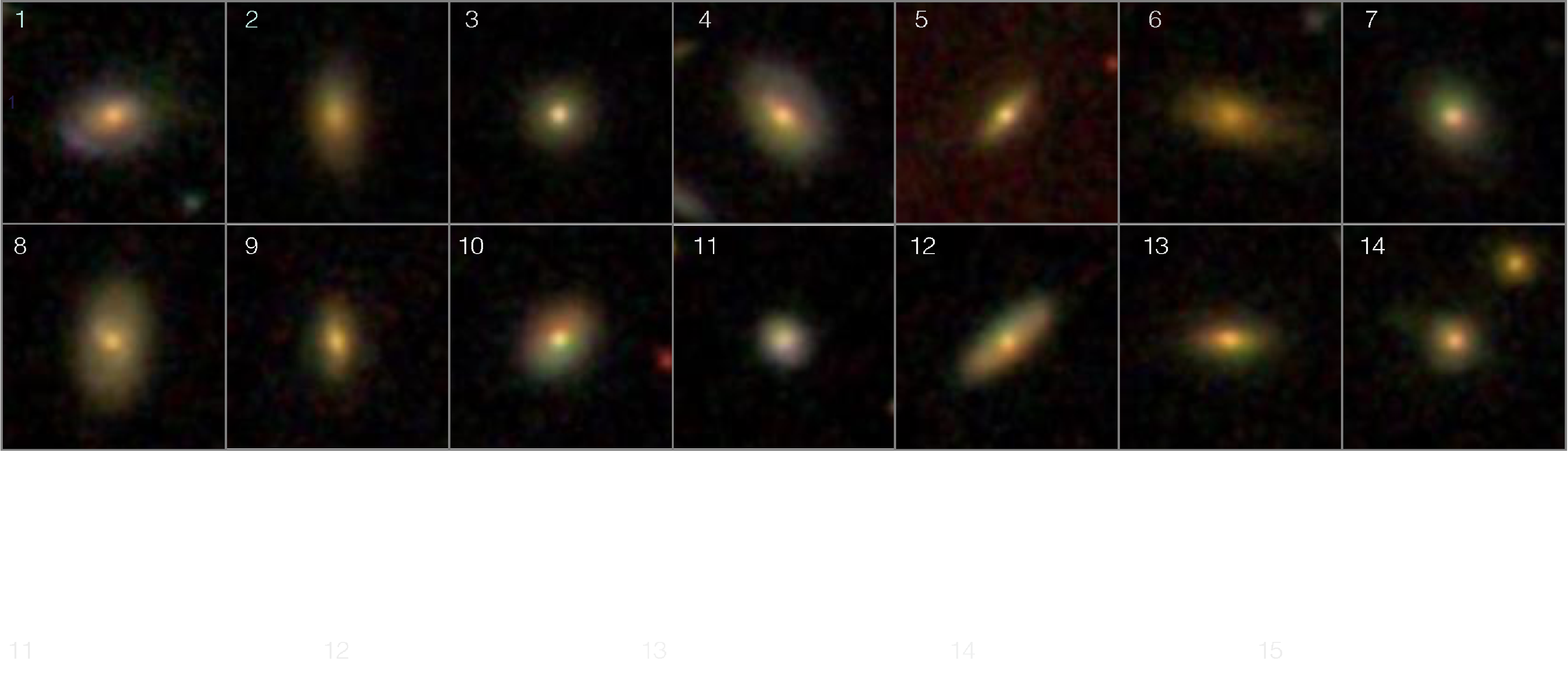}
     \caption{SDSS images of the TDE candidates. The size of image is 
      24''$\times$24''.}
    \label{images}
\end{minipage}
\end{figure*}
  
\centering
\begin{figure*}
    \centering
    \begin{minipage}[c]{\textwidth}
    \leavevmode
    \columnwidth=.45\textwidth
    \includegraphics[width={\columnwidth}]{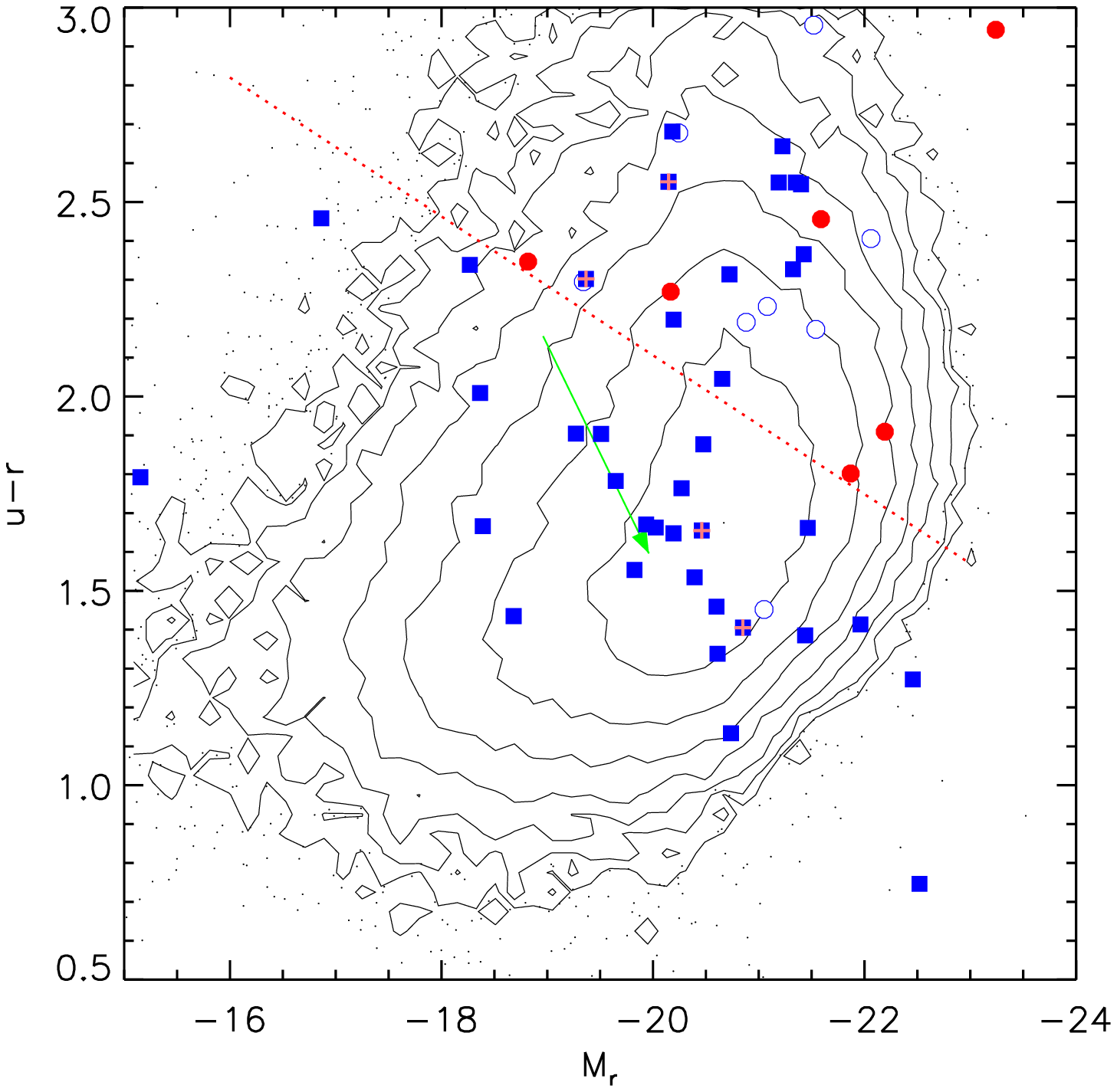}%
    \hfil
    \columnwidth=.45\textwidth
    \includegraphics[width={\columnwidth}]{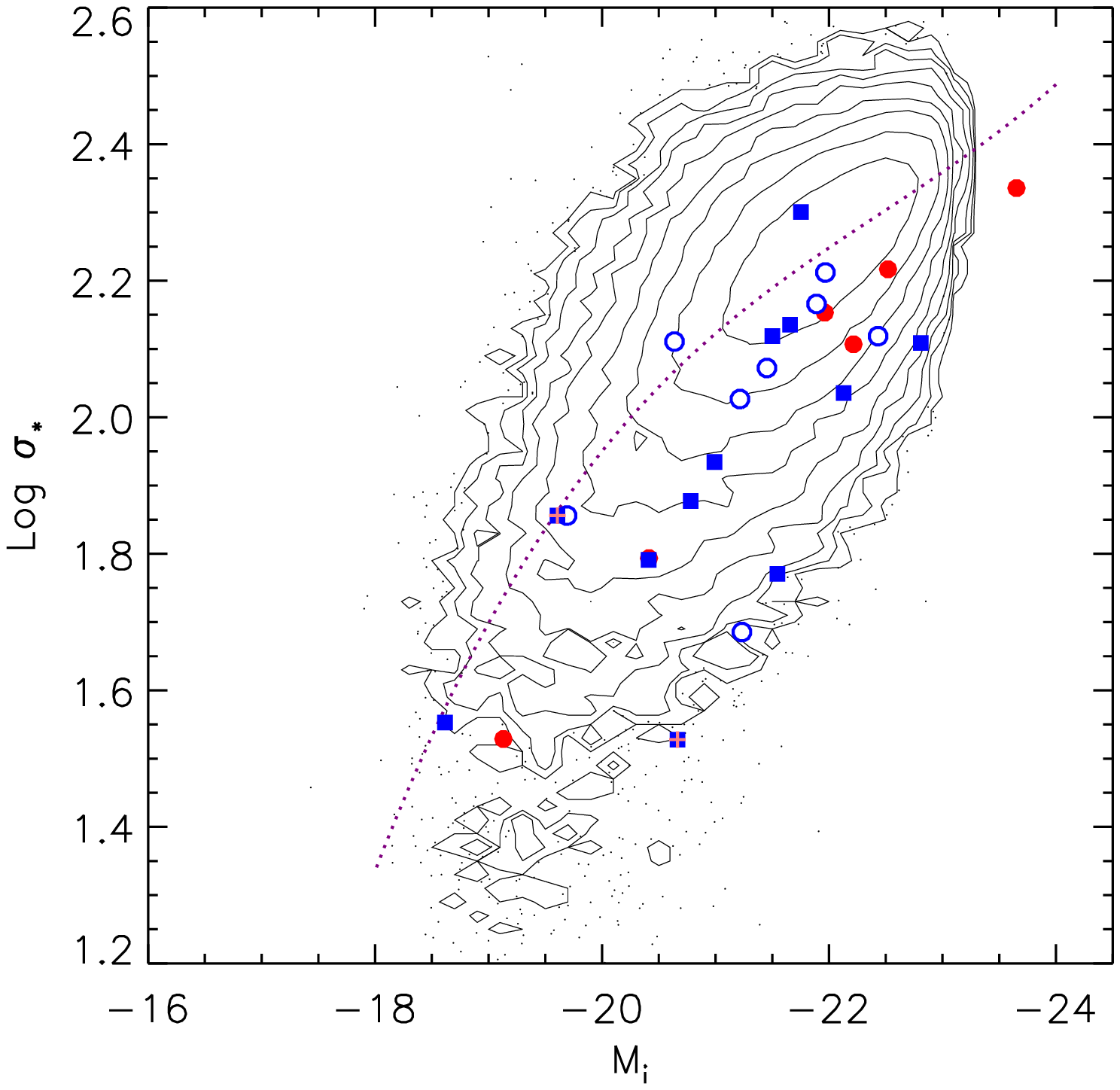}%
    \caption{The colour magnitude diagram (left panel) and velocity dispersion .vs. absolute
      magnitude (right panels). On the left panel, the contours show the density of SDSS
      spectroscopic galaxies at redshift $z<0.1$. The green arrow indicates a reddening
      correction for $A_r=1$ mag with Calzetti-type extinction curve. The red dotted line is
      an eye-guide line to define a 'red and luminous' region, where most of our objects
      locate. On the right panel, contours represent the density of early type galaxies
      in the SDSS spectroscopic sample. The purple dotted line gives the median $\sigma_*$
      at a given $M_i$ for the SDSS early type galaxies.
      All contours are given in logarithmic scale. Circles or squares represent TDE
      or TDE candidates. Filled and open circles are our sample with red for early-type galaxies
      and blue for others (see text). The squares denote previous known TDE and TDE candidates.
      Symbols with a pink cross are TDE candidates with ECLs. For J1342+0530, a TDE candidate
      with ECLs in our sample, we deliberately shift slightly the square symbol on diagram to see
      the circle.}
    \label{msigma}
    \end{minipage}
    \clearpage
\end{figure*}

\begin{figure*}
\centering
\begin{minipage}[c]{\textwidth}
  \centering
     \includegraphics[width=0.75\textwidth]{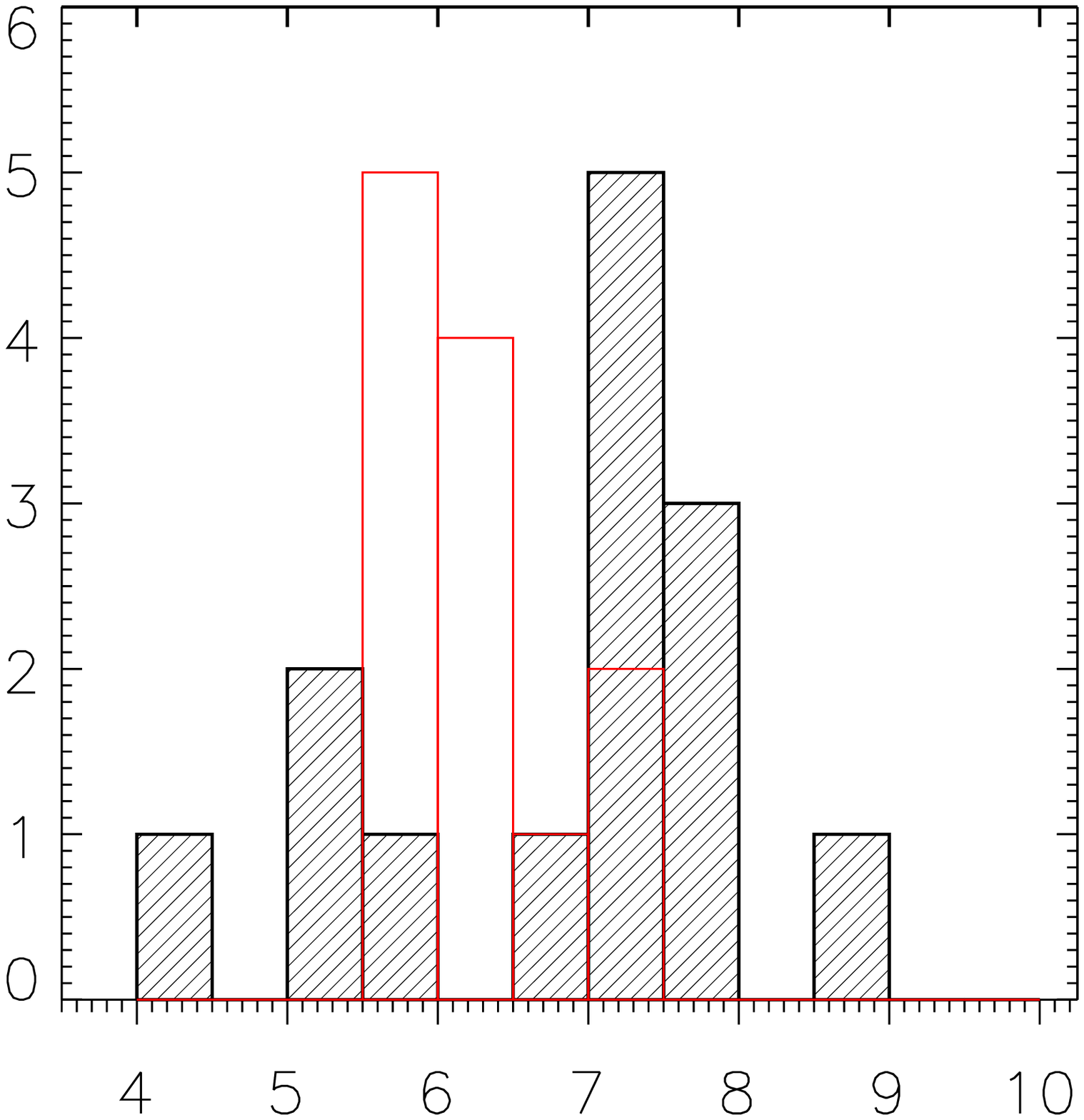}
     \vspace{0.5cm}
     \caption{A comparison of black hole mass distributions for our sample (black line) and a 
      complete optical/UV TDE host galaxies (hatched red line) by Wevers et al. (2017). Both black 
      hole masses are derived using stellar velocity dispersions. Note the BH masses in Wevers 
      et al. were estimated using the relation of Ferrarese \& Ford (2005). For consistency, we 
      transform to MM13 formula here. }
     \label{fig:mbhdist}
\end{minipage}
\end{figure*}

\begin{figure*}
\centering
\begin{minipage}[c]{\textwidth}
  \centering
     \includegraphics[width=6.0in]{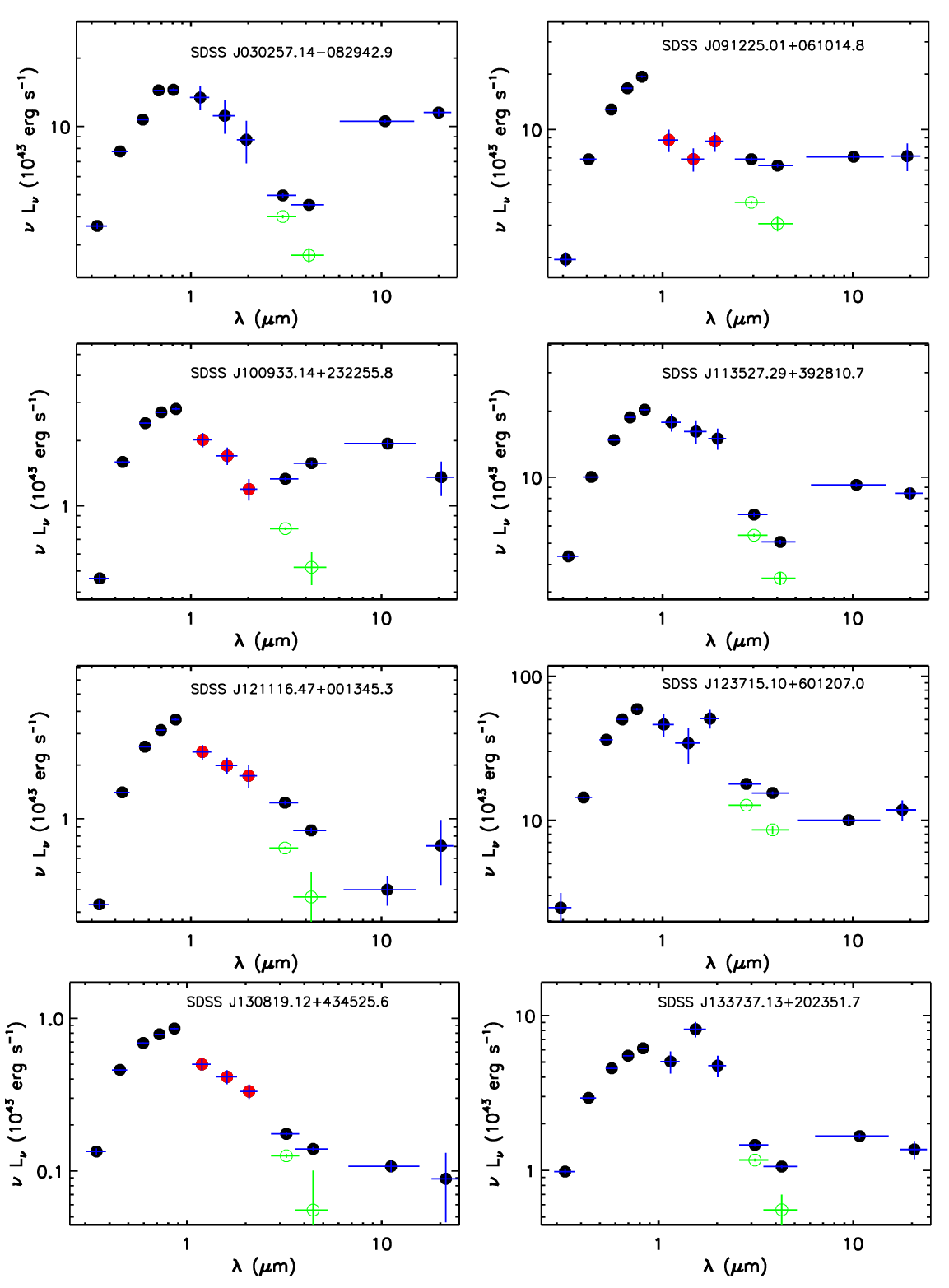}
     \caption{The spectral energy distribution from optical to mid-infrared. The 
     2MASS data are shown either as blue symbols (total magnitude of the galaxy 
     from 2MASS Extended Source Catalogue) or grey (red) symbols from Point Source 
     Catalogue. The open circles represent the observed lowest flux in W1 and W2 
     band. }
     \label{fig:sed}
\end{minipage}
\end{figure*}
\setcounter{figure}{4}

\begin{figure*}
\centering
\begin{minipage}[c]{\textwidth}
  \centering
    \includegraphics[trim=0 5cm 0 0,clip,width=6.0in]{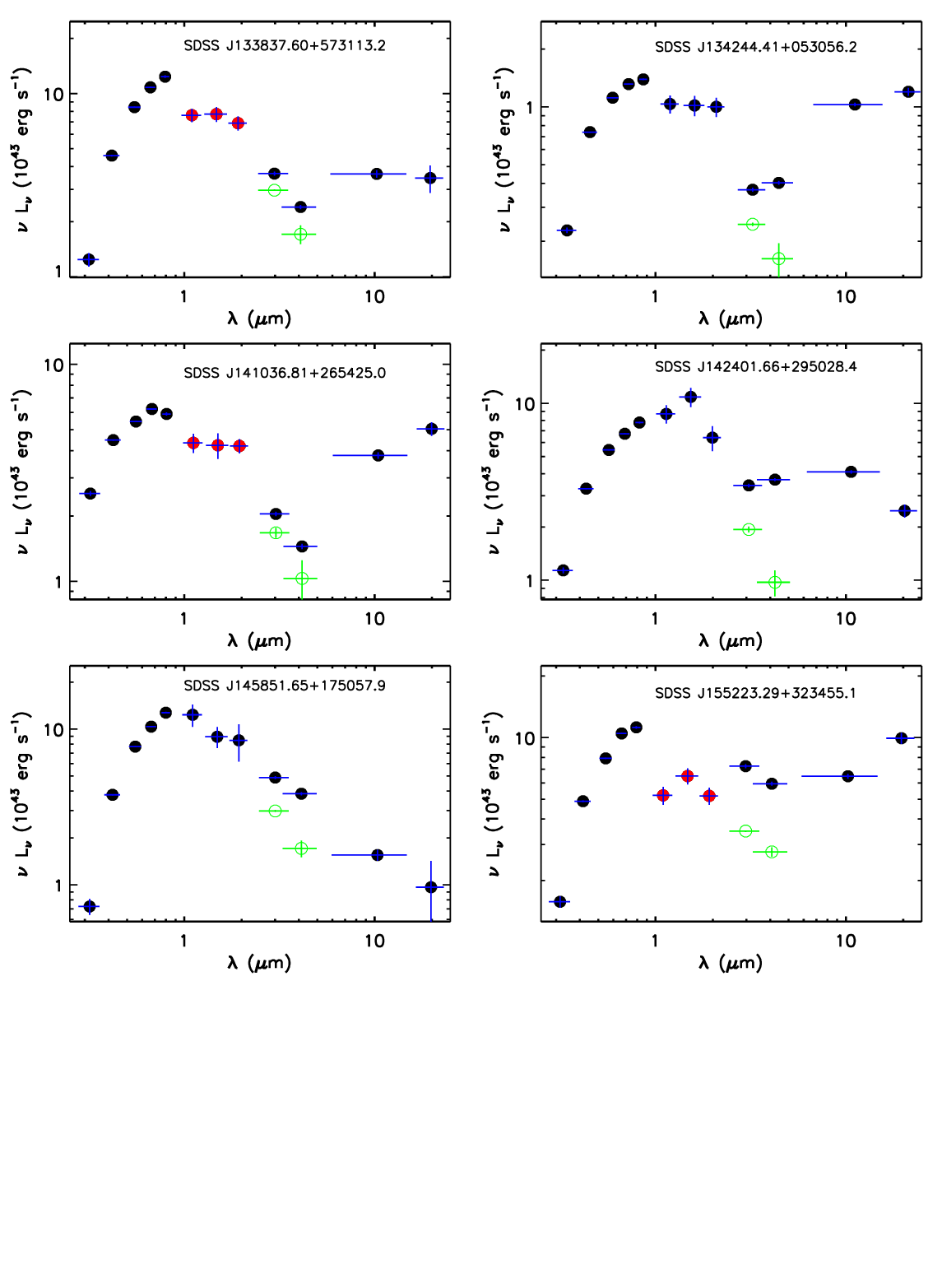}
     \caption{Continued.}
\end{minipage}
\end{figure*}

\begin{figure*}
\centering
\begin{minipage}[c]{\textwidth}
  \centering
     \includegraphics[width=0.8\textwidth]{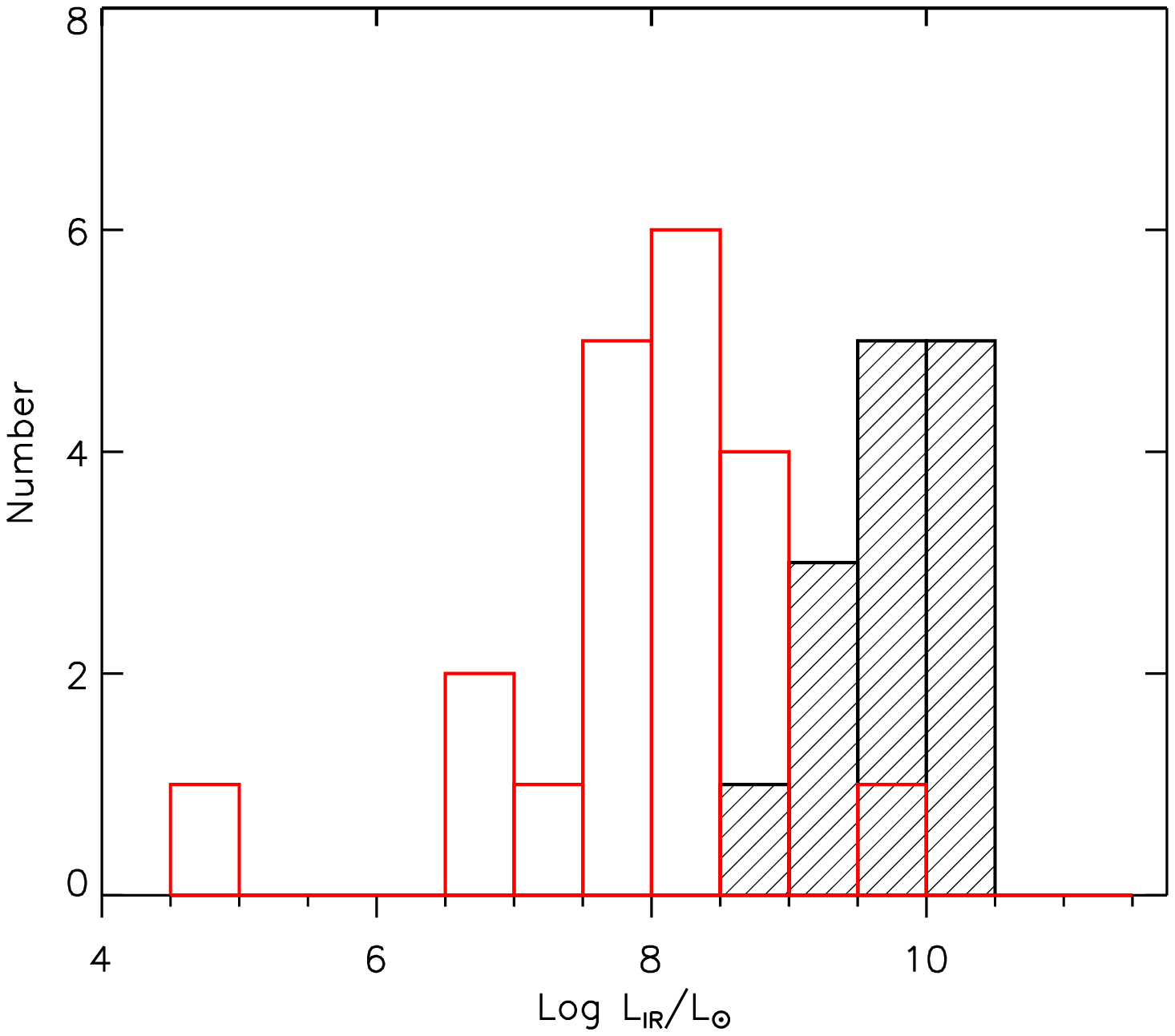}
     \caption{A comparison of galaxy subtracted MIR luminosity distribution of our sample (hatched) 
and that of type IIn supernova (red).}
     \label{lumdist}
\end{minipage}
\clearpage
\end{figure*}

\begin{figure*}
\centering
\begin{minipage}[t]{\textwidth}
  \centering
     \includegraphics[width=\textwidth]{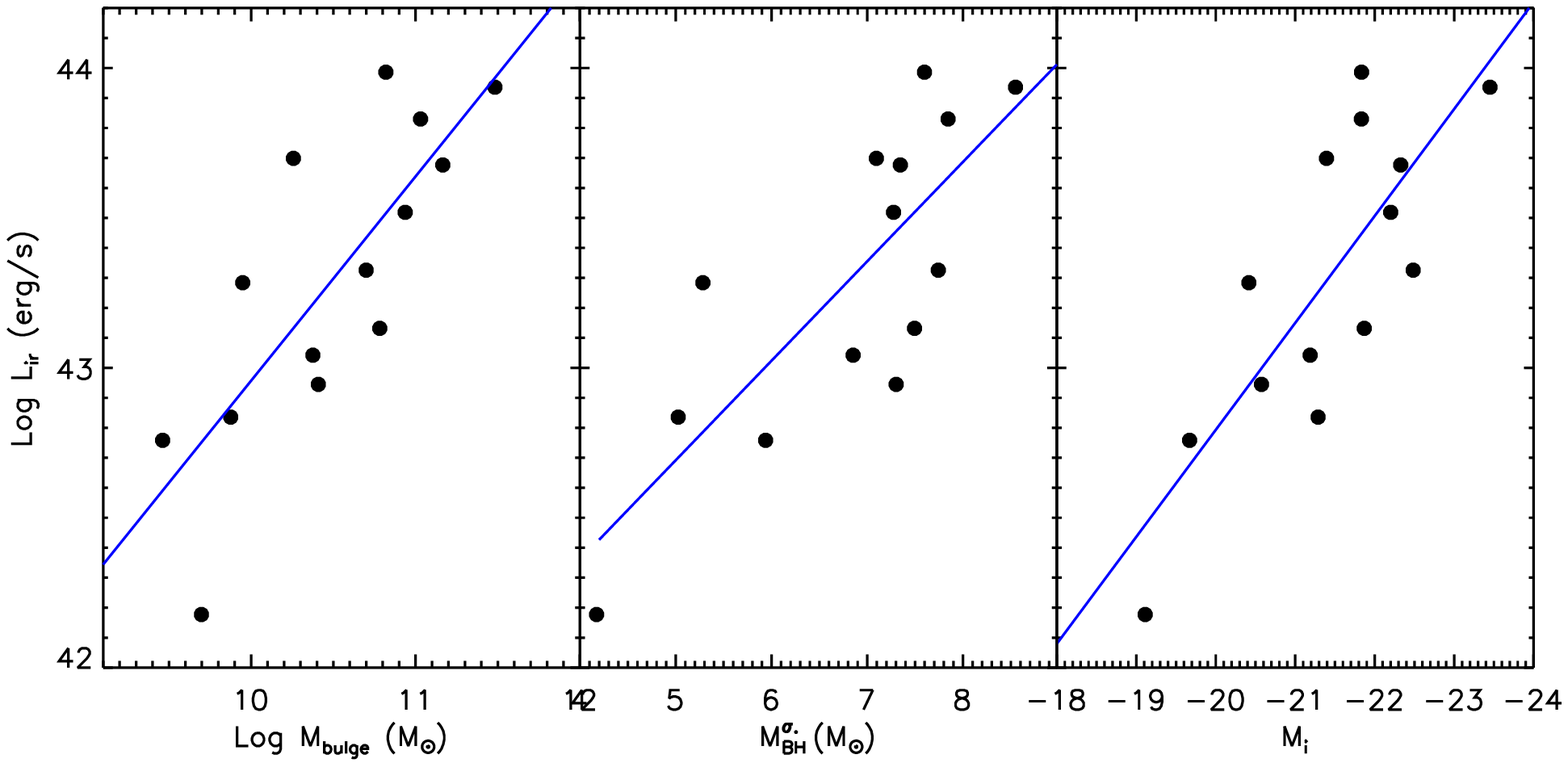}
     \caption{The correlation between the peak luminosity (black-body model)
            and the properties of the host galaxies: $L_{IR}$ versus
            $M_i^{host}$ on the left panel and $L_{IR}$ versus $\sigma_*$
            on the right panel. Blue lines show the best linear fits.}
     \label{lir_host}
\end{minipage}
\end{figure*}

\newpage
\begin{figure*}
\centering
 \begin{minipage}[t]{\textwidth}
  \centering
     \includegraphics[width=0.8\textwidth]{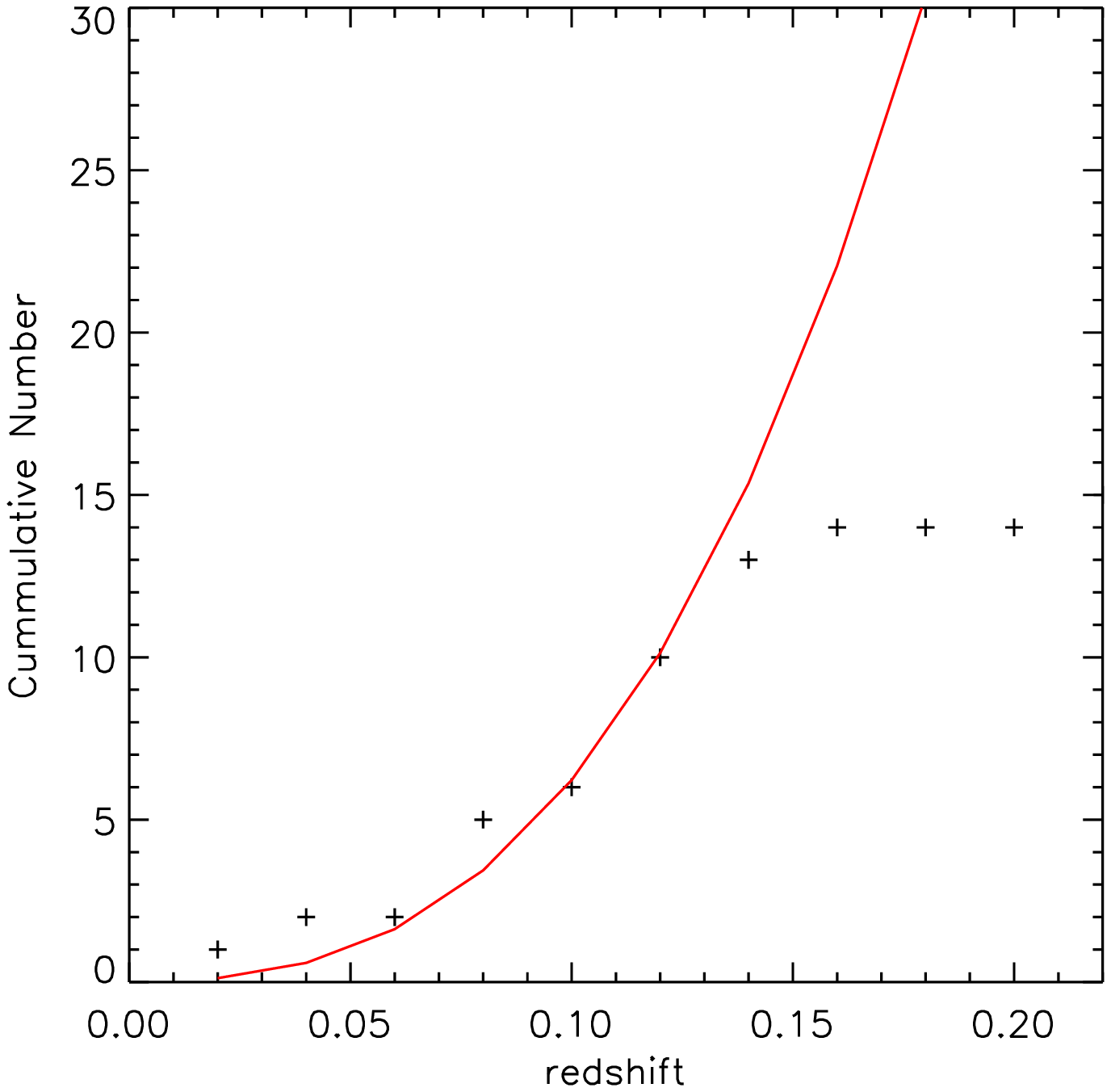}
     \caption{The accumulated redshift distribution of the sample. The curve shows
        the comoving volume of SDSS sky at the correspondent redshift
        multiplied $8.3\times 10^{-8}$ Mpc$^3$.}
     \label{redshift}
\end{minipage}
\label{lastpage}
\end{figure*}

    \clearpage
    \thispagestyle{empty}
    \begin{landscape}
    \begin{table} 
    \centering
  \captionof{table}{The sample of Spectroscopically normal galaxies with variable infrared emission.}
  \begin{tabular}{@{}rlrrrrlrrrrrrrrr@{}}
  \hline
No & IAU & z & $M_i$ & $g-r$ & type & $W(H\alpha)$ & $\sigma_*$  & $B/T$ & $\log M^{bul}$ &W1 & W2 & W3 & W4 & V &  $\delta V^u$ \\
   &     &   & mag   &       &      &   \AA        &   km\ s$^{-1}$ &   & M$_{\sun}$ &mag & mag & mag & mag  & mag & mag \\
  & (1) & (2) & (3) & (4) & (5) & (6) & (7) & (8) & (9) & (10) & (11) & (12) & (13) & (14) & (15) \\
\hline
  1&  030257.14-082942.9&0.1056&-22.20&  0.65& LINER&  83.9$\pm$2.9& 127$\pm$13& 0.59$\pm$0.04&10.94$_{-0.13}^{+0.10}$&13.09&12.21& 8.45& 6.21 & 16.744 & 0.011\\
 2&  091225.01+061014.8&0.1453&-22.33&  0.98& SF&  13.1$\pm$0.9& 131$\pm$15& 0.52$\pm$0.05&11.17$_{-0.14}^{+0.10}$&13.52&12.63& 9.67& 7.51 & 17.479 & 0.014 \\
 3&  100933.14+232255.8&0.0719&-20.42&  0.75& SF&   2.0$\pm$0.2&  54$\pm$6& 0.58$\pm$0.03& 9.95$_{-0.14}^{+0.12}$&13.60&12.44& 9.38& 7.61 & 17.431 & 0.036\\
 4&  113527.29+392810.7&0.1084&-22.48&  0.72& LINER&  45.9$\pm$1.5& 155$\pm$10& 0.32$\pm$0.02&10.70$_{-0.17}^{+0.12}$&12.82&12.15& 8.66& 6.61 & 16.581 & 0.031\\
 5&  121116.47+001345.3&0.0763&-20.58&  0.94& AbsL &   0.4$\pm$0.2& 129$\pm$8& 0.79$\pm$0.05&10.41$_{-0.14}^{+0.10}$&13.83&13.23&11.23& 8.46 & 17.588 & 0.062\\
 6&  123715.10+601207.0&0.2164&-23.45&  1.30& SF&   3.3$\pm$1.0& 220$\pm$17& 0.46$\pm$0.04&11.48$_{-0.14}^{+0.10}$&13.51&12.68&10.32& 7.99 & 17.532 & 0.061\\
 7&  130819.12+434525.6&0.0365&-19.11&  0.74& SF&   6.5$\pm$0.4&  33$\pm$27& 0.99$\pm$0.03& 9.70$_{-0.16}^{+0.11}$&14.24&13.51&10.96& 9.01 & 17.282 & 0.010\\
 8&  133737.13+202351.7&0.0725&-21.19&  0.78& LINER&   9.9$\pm$0.4& 106$\pm$7& 0.32$\pm$0.02&10.38$_{-0.13}^{+0.10}$&13.52&12.88& 9.56& 7.62 & 16.678 & 0.027\\
9&  133837.60+573113.2&0.1269&-21.87&  0.96& LINER&  13.6$\pm$0.6& 140$\pm$10& 0.67$\pm$0.04&10.78$_{-0.14}^{+0.11}$&13.87&13.34&10.06& 7.97 & 17.575 & 0.040\\
10&  134244.41+053056.2&0.0365&-19.67&  0.75& SF&  15.9$\pm$0.4&  71$\pm$5& 0.34$\pm$0.07& 9.46$_{-0.20}^{+0.13}$&13.43&12.35& 8.50& 6.19 & 16.817 & 0.011\\
11&  141036.81+265425.0&0.1071&-21.29&  0.51& LINER&  60.6$\pm$2.1&  48$\pm$11& 0.09$\pm$0.05& 9.88$_{-0.22}^{+0.20}$&14.09&13.48& 9.60& 7.14 & 17.459 & 0.021\\
12&  142401.66+295028.4&0.0855&-21.39&  0.84& LINER&  13.5$\pm$0.5& 118$\pm$9& 0.13$\pm$0.03&10.26$_{-0.17}^{+0.14}$&12.98&11.92& 8.97& 7.37 & 16.944 & 0.019\\
13&  145851.65+175057.9&0.1165&-21.83&  1.07& SF&   0.9$\pm$0.3& 162$\pm$11& 0.69$\pm$0.05&11.03$_{-0.15}^{+0.10}$&13.35&12.62&10.77& 9.14 & 17.348 & 0.020\\ 
14&  155223.29+323455.1&0.1277&-21.83&  0.82& LINER&  25.8$\pm$1.0& 146$\pm$21& 0.81$\pm$0.04&10.82$_{-0.16}^{+0.12}$&13.14&12.38& 9.45& 6.83 & 17.451 & 0.040\\
\hline \\
\end{tabular}
\footnotesize{Column (1): IAU name;  (2): redshift; (3) absolute magnitude at $i$; (4) SDSS colour $g-r$;  
(5) spectral classification based on BPT diagram: SF-star forming galaxy, AbsL-absorption line galaxy; (6) equivalent width of H$\alpha$; (7) stellar velocity 
dispersions; (8) bulge to total light ratio in $r$ band from  \citet{Simard11};  (9) bulge stellar mass 
from \citet{Mendel14} ; (10-13) ALLWISE magnitudes from W1 to W4 bands; (14-15) are the mean and standard deviations of quarter median of CRTS  
magnitude in V. }
\label{table1}
\end{table} 
\end{landscape}
\clearpage

\clearpage
\thispagestyle{empty}
\begin{landscape}
\begin{table} 
\centering 
\captionof{table}{Results of MIR Light Curve Analysis.}
\begin{tabular}{@{}rccccccccccccccccc@{}}
  \hline
\multirow{2}{*}{No} & \multirow{2}{*}{$W1^{gal}$}& \multirow{2}{*}{$W2^{gal}$}& \multirow{2}{*}{$\Delta W1$} & \multirow{2}{*}{$\Delta W2$} &
\multirow{2}{*}{$\tau_{W1}$}& \multirow{2}{*}{$\tau_{W2}$} & \multicolumn{2}{c}{black body}& \multicolumn{3}{c}{0.01$\mu$m silcate} & \multicolumn{3}{c}{0.1$\mu$m silcate} & \multicolumn{3}{c}{1$\mu$m silcate}\\
&  & & & & & &   $\log L_{IR}$ & $T_{BB}$ &
$\log L_{IR}$ & $T_d$ & $\log M_d$ & $\log L_{IR}$ & $T_d$ & $\log M_d$ & $\log L_{IR}$ & $T_d$ & $\log M_d$ \\ 
 & mag & mag & mag & mag & yr  & yr & erg~s$^{-1}$ & K &  erg~s$^{-1}$ & K & $M_{\sun}$ & erg~s$^{-1}$ & K & $M_{\sun}$ & erg~s$^{-1}$ & K & $M_{\sun}$ \\ 
 & (1) & (2) & (3) & (4) & (5) & (6) & (7) & (8) & (9) & (10) & (11) & (12) & (13) & (14) & (15) & (16) & (17) \\ \hline
 1&13.67$\pm$0.08&13.01$\pm$0.21& 0.36& 0.72& 0.81& 1.28&43.52& 632$\pm$9&43.92& 561$\pm$7& 0.263&43.69& 553$\pm$6& 0.047&43.51& 513$\pm$6&-0.095\\
 2&13.95$\pm$0.29&13.18$\pm$0.21& 0.73& 0.95& 0.51& 0.71&43.68& 776$\pm$22&43.97& 670$\pm$16& 0.069&43.73& 659$\pm$15&-0.146&43.56& 603$\pm$13 &-0.285\\
 3&14.79$\pm$0.14&14.39$\pm$0.20& 0.85& 1.46& 0.48& 0.75&43.28& 637$\pm$10&43.68& 565$\pm$8& 0.012&43.45& 557$\pm$7&-0.204&43.27& 517$\pm$6 &-0.346\\
 4&13.35$\pm$0.31&12.93$\pm$0.35& 0.31& 0.49& 0.60& 0.99&43.33& 777$\pm$22&43.62& 670$\pm$16&-0.282&43.38& 659$\pm$15&-0.497&43.21& 604$\pm$13 &-0.637\\
 5&14.19$\pm$0.19&13.68$\pm$0.18& 0.82& 1.12& 0.71& 1.29&42.94&1041$\pm$72&43.07& 853$\pm$46&-1.154&42.83& 834$\pm$44&-1.367&42.68& 745$\pm$35 &-1.500\\
 6&13.90$\pm$0.46&13.26$\pm$0.65& 0.45& 0.73& 0.77& 1.06&43.94& 749$\pm$10&44.25& 650$\pm$7& 0.390&44.01& 639$\pm$7& 0.174&43.84& 587$\pm$6& 0.034\\
 7&13.87$\pm$0.31&13.63$\pm$0.41& 0.50& 1.22& 0.31& 0.45&42.17& 667$\pm$13&42.55& 588$\pm$10&-1.175&42.32& 579$\pm$9&-1.390&42.14& 536$\pm$8 &-1.532\\
8&13.40$\pm$0.16&12.98$\pm$0.22& 0.37& 0.98& 0.70& 0.77&43.04& 605$\pm$8&43.47& 540$\pm$6&-0.140&43.23& 533$\pm$6&-0.356&43.05& 496$\pm$5&-0.499\\
9&14.17$\pm$0.20&13.82$\pm$0.20& 0.41& 0.58& 0.25& 0.33&43.14& 985$\pm$34&43.30& 816$\pm$23&-0.866&43.06& 800$\pm$22&-1.080&42.91& 718$\pm$17 &-1.214\\
10&13.65$\pm$0.26&13.38$\pm$0.30& 0.53& 1.16& 3.08& 3.20&42.76& 569$\pm$31&43.21& 511$\pm$25&-0.318&42.98& 504$\pm$24&-0.534&42.79& 471$\pm$21 &-0.676\\
11&14.72$\pm$0.15&14.12$\pm$0.16& 0.42& 0.66& 0.08& 0.08&42.84& 938$\pm$92&43.02& 784$\pm$63&-1.089&42.78& 769$\pm$60&-1.303&42.63& 693$\pm$48 &-1.439\\
12&13.60$\pm$0.28&13.16$\pm$0.36& 0.72& 1.64& 1.74& 1.59&43.70& 597$\pm$18&44.13& 533$\pm$14& 0.539&43.90& 526$\pm$13& 0.323&43.71& 491$\pm$12 
&0.181\\
13&14.09$\pm$0.24&13.67$\pm$0.41& 0.66& 1.01&12.58& 6.43&43.83&1279$\pm$***&43.84& 999$\pm$***&-0.607&43.60& 973$\pm$***&-0.819&43.46& 852$\pm$*** &-0.945\\
14&14.13$\pm$0.30&13.31$\pm$0.22& 0.94& 0.96& 7.18& 7.92&43.99&1145$\pm$388&44.06& 919$\pm$230&-0.270&43.82& 897$\pm$218&-0.482&43.67& 794$\pm$166 &-0.612\\ \hline 
\end{tabular}
\footnotesize{Column (1) and (2): the estimate of constant background magnitudes in $W1$ and $W2$ 
bands from the light curve fitting; Column (3) and (4): the variability amplitude in $W1$ and $W2$ 
bands; Column (5) and (6): the decay time of exponential law fit to the light curves in $W1$ and $W2$ 
bands; Column (7)-(8) luminosity and temperature of black body model derived from the peak flux of 
the flare;  Column (9)-(11), (12)-(14) and (15)-(17): logarithmic luminosity, temperature and logarithmic 
dust mass for one temperature silicate dust model with grain sizes 0.01, 0.1 and 1 microns.} 
\label{table2}
\end{table}
\end{landscape}
\clearpage

\appendix
\section{Light curves and spectra of different spectroscopic classes}

In the main text, we focused on the subsample of non-Seyfert galaxies with 
flare-like light curves, and their statistical properties are compared with 
spectroscopically classified Seyfert galaxies and blazars. In order to for readers 
to gain a direct view of the difference between different classes, in this appendix, 
we present the MIR light curves and SDSS spectra for the leading 7 objects (in 
increasing RA order) in the subsamples of Seyfert 1 galaxies, Seyfert 2 
galaxies, Blazars and non-Seyfert galaxies that are not included in Figure 
\ref{lcsp}. 
\renewcommand\thefigure{\thesection\arabic{figure}} 
\setcounter{figure}{0} 
\clearpage
\thispagestyle{empty}
\begin{figure*}
\centering
\begin{minipage}[c]{\textwidth}
  \centering
     \includegraphics[width=6.0in]{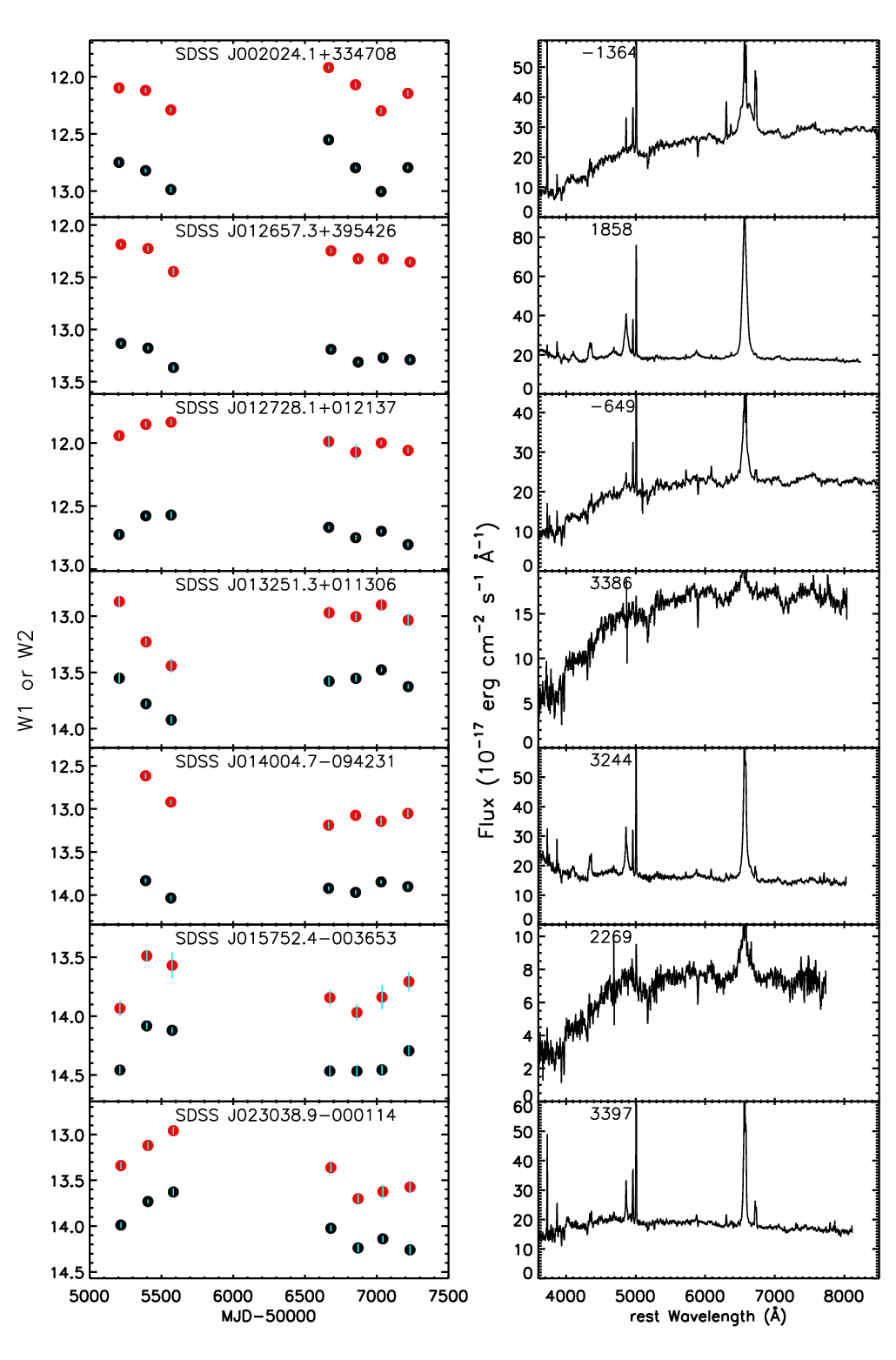}
     \caption{The SDSS spectra (right panels) and WISE light curves (left panels) 
     of sources Seyfert 1 galaxies. Left panels: The light curves in W1 and W2 are 
     represented with black and red circles.  Right panels: the number on the upper 
     corner gives the time gap between the first WISE observation and the latest 
     SDSS spectroscopic observation; a negative value means that spectrum was taken 
     before WISE observation. When there is/are more than one spectrum, we plot them
     in a light colour. }
     \label{sey1lcsp}
\end{minipage}
\end{figure*}
\thispagestyle{empty}
\begin{figure*}
\centering
\begin{minipage}[c]{\textwidth}
  \centering
     \includegraphics[width=6.0in]{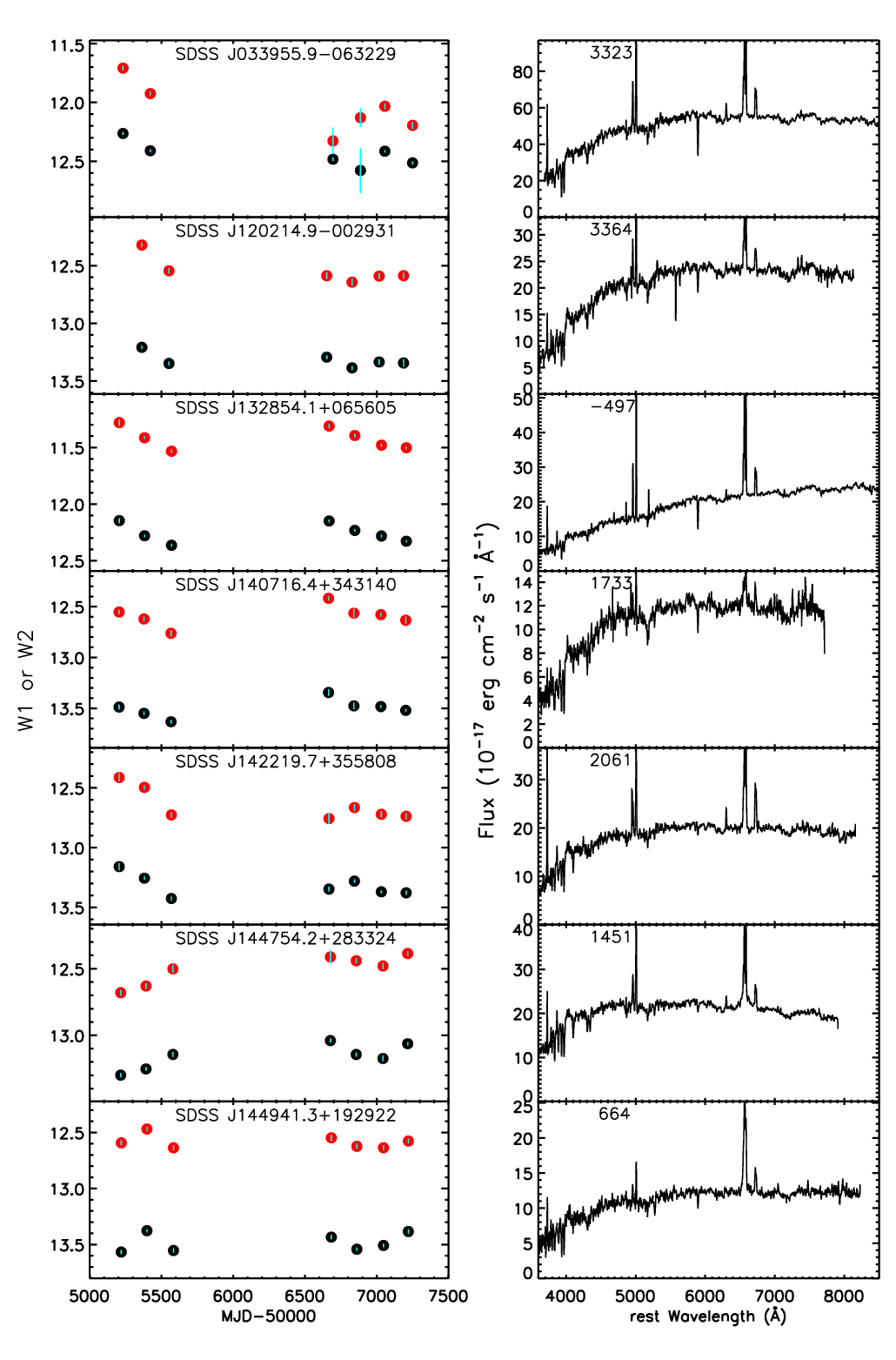}
     \caption{Same as Figure A1, but for Seyfert 2 galaxies.}
     \label{sey2lcsp}
\end{minipage}
\end{figure*}
\thispagestyle{empty}
\begin{figure*}
\centering
\begin{minipage}[c]{\textwidth}
  \centering
     \includegraphics[width=6.0in]{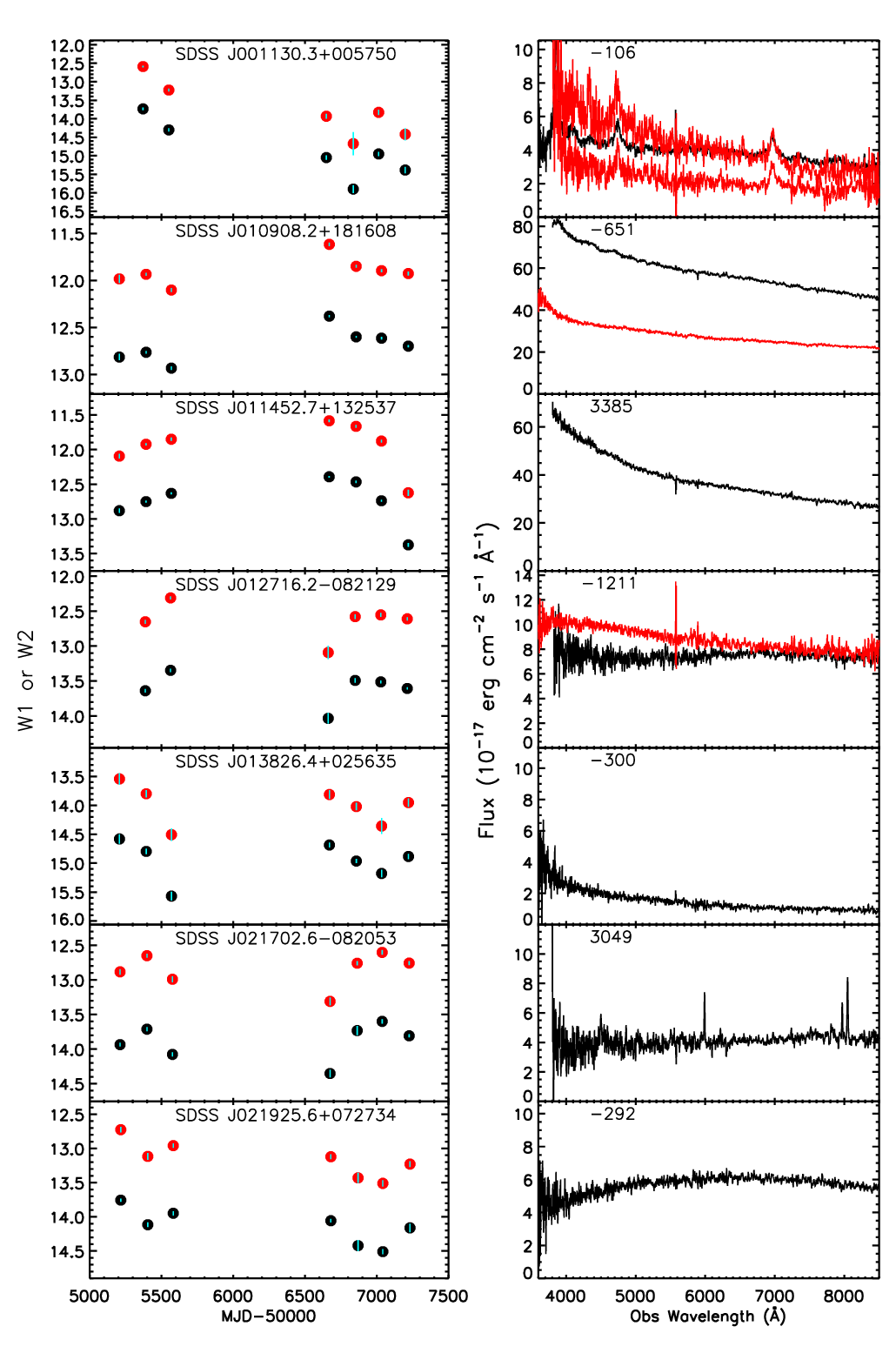}
     \caption{Same as Figure A1, but for Blazars. Note now the wavelengths are 
     in the observer's frame due to lack of reshifts for some objects.}
     \label{blzlcsp}
\end{minipage}
\end{figure*}
\thispagestyle{empty}
\begin{figure*}
\centering
\begin{minipage}[c]{\textwidth}
  \centering
     \includegraphics[width=6.0in]{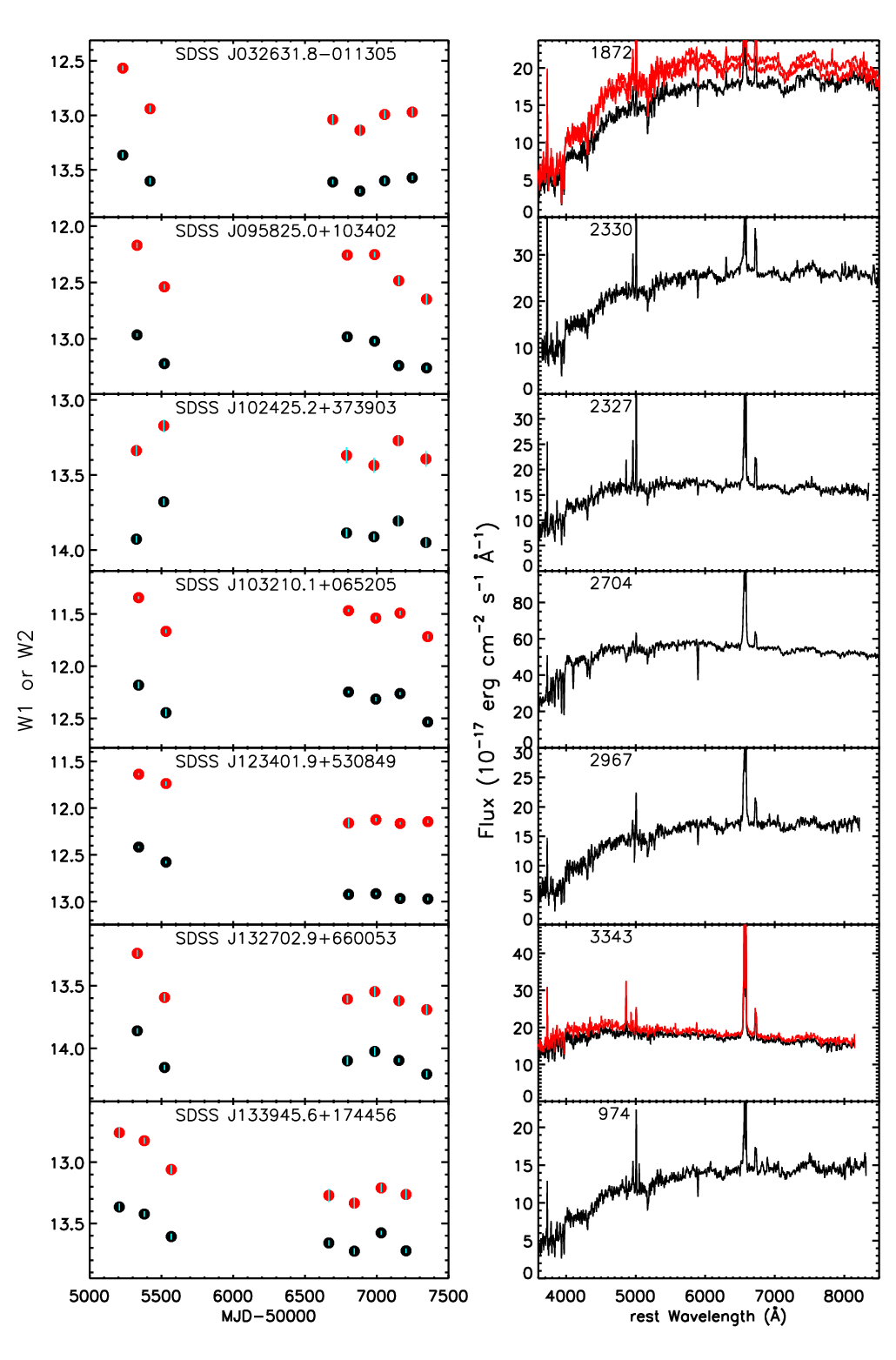}
     \caption{Same as Figure A1, but for some non-Seyfert galaxies not in Figure \ref{lcsp}.}
     \label{gallcsp}
\end{minipage}
\end{figure*}

\end{document}